\documentclass[a4paper,11pt]{article}
\usepackage{amsfonts}

\usepackage{graphicx}
\graphicspath{{Figures/}}
\usepackage{amsmath}
\usepackage{amssymb}
\usepackage{latexsym}
\usepackage[colorlinks]{hyperref}
\usepackage{color}
\usepackage{float}
\usepackage{cite}
\usepackage{makeidx}
\usepackage{colortbl}
\usepackage{csquotes}
\usepackage[squaren]{SIunits}

\usepackage[textfont={footnotesize,sf},labelfont={color=blue,bf,sf},labelsep=endash]{caption}
\usepackage[position=top,labelfont={color=blue,bf,sf}]{subfig}
\usepackage{bm}

\newcommand{\bra}{\begin{array}}
\newcommand{\era}{\end{array}}
\newcommand{\beq}{\begin{equation}}
\newcommand{\eeq}{\end{equation}}
\newcommand{\bqr}{\begin{eqnarray}}
\newcommand{\eqr}{\end{eqnarray}}

\def\BC{\bb C}
\def\_\BC{\bbi C}

\def\( {\left(}
\def\) {\right)}
\def\no2 {{\textstyle{n\over 2}}}
\def \const {{\rm constant}}

\newcommand{\lga}{\longrightarrow}

\begin{document}
\begin{titlepage}
\setcounter{page}{1}
\renewcommand{\thefootnote}{\fnsymbol{footnote}}
\begin{flushright}
\end{flushright}
\vspace{5mm}
\begin{center}
{\Large \bf {   
Classical Instability Effects on Photon Excitations\\  and Entanglement
}}\\

\vspace{5mm}

{\bf Radouan Hab-arrih}$^{a}$,   
{\bf Ahmed Jellal\footnote{\sf a.jellal@ucd.ac.ma}}$^{a,b}$ and  
{\bf Abdeldjalil Merdaci}$^{c}$

\vspace{5mm}
{$^{a}$\em Laboratory of Theoretical Physics,  
Faculty of Sciences, Choua\"ib Doukkali University},\\
{\em PO Box 20, 24000 El Jadida, Morocco}

{$^{b}$\em Canadian Quantum  Research Center,
	204-3002 32 Ave Vernon, \\ BC V1T 2L7,  Canada}

{$^{c}$\em Facult\'e des Sciences, Universit\'e 20 Ao\^ut 1955 Skikda,\\ BP 26, Route El-Hadaiek 21000, Algeria}

\vspace{30mm}

\begin{abstract}
  The Schr\"{o}dinger dynamics of photon excitation numbers  together with entanglement in two non-resonant time-dependent coupled oscillators is investigated. By considering 
  $ \pi-$periodically pumped
  parameters and using suitable transformations, 
   we obtain the coupled Meissner oscillators. 
  Consequently, our analytical study 
     shows two interesting results, which can be summarized as follows. (i):  Classical instability of  classical analog of quantum oscillators and photon excitation {averages  $\left\langle N_{j}\right\rangle  $} are strongly correlated.  (ii): Photon excitation's and entanglement are connected to each other. These  results
  can be used to shed light on the link between quantum systems and their classical counterparts. Also it allow to  control  entanglement by engineering  only   classical systems where the experiments are less expensive.
\vspace{30mm}

\noindent {\bf PACS numbers}: 03.65.Fd, 03.65.Ge, 03.65.Ud, 03.67.Hk
\\  
\noindent {\bf Keywords}: Classical instability maps, time-dependent coupled oscillators, photon excitation's, entanglement, Ermakov equation, Meissner equation.
\end{abstract}
\end{center}
\end{titlepage}

\section{Introduction}

Since the 
emergence of quantum theory (QT) in the beginning of $ 20 ^{th} $ century, entanglement was used to refute the basic QT's principles. In the early stages, Einstein, Podolsky and Rosen (EPR) 
\cite{R01} (known as EPR paper), have attacked violently QT by remarking that wave functions can be  entangled, which entails in their point of view the existence of hidden variables. In other part, entanglement was considered as a necessary complement of QT because  without it, it is impossible to interpret and confirm the previsions of QT \cite{R02}. Actually, entanglement plays  an important role in quantum information processing protocols and it is considered a necessary resource to go beyond the classical communications and technologies. 

In the  last years,  controlling  entanglement in  time-dependent coupled harmonic oscillators was extensively studied, especially when oscillator systems  in contact with environment. For instance,  
It was shown that
the possibility to generate entanglement by phasing control in two \cite{R03}
and  three \cite{R04}  isotropic harmonic oscillator's sinusoidally coupled to each other by $c(t)=c_{0}\cos(\omega t)$    and  weekly coupled to   an  harmonic bath. 
It was found that the survival of entanglement for a large simulation time is due to instability of decoupled (from the bath) normal oscillator \cite{R05}. More recently,  it was shown that the vacuum $ |G\rangle $  of two time-independent resonant oscillators contains virtual excitation's \cite{R7}, i.e. $\langle G|a^{+}a |G\rangle=\langle G|b^{+}b |G\rangle\neq 0$, in the range of strong coupling,
which a consequence of the {counter-rotating  terms} appearing in the Hamiltonian.
A a result, the presence of these  excitation's  maintains entanglement between oscillators. 

Motivated by  the above studies, we address to the question: 
how  classical instabilities affect photon excitation's and consequently entanglement in the vacuum of two time-dependent non-resonant coupled harmonic oscillators. 
Our response will be given in the framework of 
an  assumption 
based on the fact that
our harmonic oscillators are connected by a periodically quenched coupling parameter $J(t)=J_{0} \Theta (t) $ and  having perturbed frequencies 
{$\omega_{1,2}^{2}(t)=\omega_{0}^{2}\pm \epsilon \Theta(t)$}, with   $ \Theta $  is a $ \pi $-periodic function $\Theta(t)=\Theta(t+\pi)$ and $ \epsilon $ is the quench amplitude. As a result, 
we end up with an integrable 
model called  {two coupled} Meissner oscillators \cite{R06}.
{This kind of oscillators can be seen, for instance, as    $ LC_{0} $    oscillator with the parametric frequency $ \omega_{0}^{2}=(LC_{0})^{-1} $ where  the capacitance $ C_{0} $ is pumped by a voltage $ V(t) $ such that $C_{0}\longrightarrow C(t)=C_{0}+ C_{p}\Theta(t) $ $ (C_{p}<C_{0}) $ \cite{R10}, or as  a charged pendulum
in alternating, piece-wise constant, homogeneous electric field \cite{R15}}.
The resolution of {the Schr\"{o}dinger} dynamics allows us to find Ermakov equations \cite{R9} and utilization of  suitable transformations leads to get two Meissner differential equations  corresponding to classical counterparts of the decoupled Hamiltonian. Then, we study the instabilities of derived differential equations and show  instability/stability {diagrams}. With these, 
we be able to  investigate the link between two strongly different features: 
the photon excitation's and classical instabilities. Additionally, by computing logarithmic negativity we establish a bijection   between entanglement and photon excitation's.

The layout of our paper is given as follows. In {\color{blue}Sec.} \ref{sec2}, we present our model and  show how to  exactly decouple the  Hamiltonian system by using suitable transformations.  The instabilities of emerged equations of both Ermakov and Meissner will be discussed in {\color{blue}Sec.} \ref{Sec3}.  We compute entanglement by using logarithmic negativity and quantifying excitation in both oscillators by averaging the number operators over the vacuum in {\color{blue}Sec.}~\ref{sec4}. We show our numerical results and present different discussions in {\color{blue}Sec.} \ref{sec5}. Finally, we give an exhaustive conclusion to our work. 

\section{Model and Schr\"{o}dinger dynamics \label{sec2}}
\subsection{Model and integrability }

\begin{figure}[htbphtbp]
	\centering
\includegraphics[height=6cm, width=8cm]{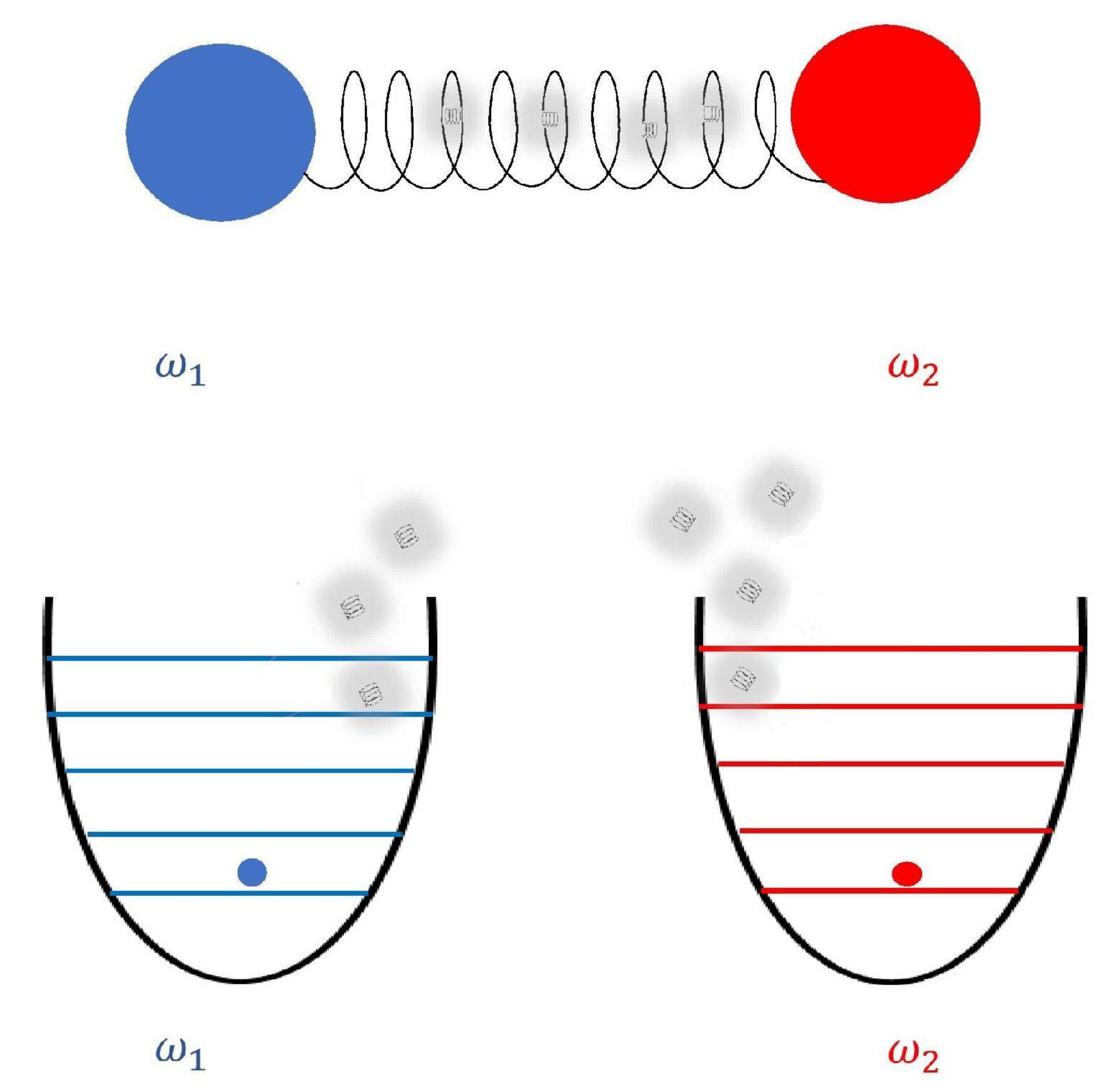}
	\captionof{figure}{\sf (color online)  The schematic shows two coupled oscillators via a position-position coupling type $ x_{1}x_{2} $. The particles still in their vacuum. The dynamics generates virtual excitation's between oscillators that affect quantum quantities.} 
	\label{f1}   
\end{figure}

The main concern in the present work is to answer the question asked in our introduction. Mainly about how
the  classical instabilities affect photon excitation's and therefore entanglement in the vacuum of two time-dependent non-resonant coupled harmonic oscillators ({\color{blue}Figure }\ref{f1})
described by the Hamiltonian
%
\begin{eqnarray}
H(\hat{x}_{1},\hat{x}_{2})=\frac{\hat{p}_{1}^{2}}{2}+\frac{\hat{p}_{2}^{2}}{2}+\frac{1}{2}\omega_{1}^{2}(t)\hat{x}_{1}^{2}+\frac{1}{2}\omega_{2}^{2}(t)\hat{x}_{2}^{2}-J(t)\hat{x}_{1} \hat{x}_{2}\label{ham}
\end{eqnarray} 
where $\omega_{j}(t)$  are the frequencies and  $ J(t) $ is a coupling parameter, 
with $ j=1,2 $. For simplicity, we assume that the masses are unit (we set the masses as
equal to one, because, as Macedo and Guedes showed \cite{R2}, a simple transformation may be applied that
makes the assumption valid) and $\hbar=1 $.
 
 Since the Hamiltonian \eqref{ham} is involving an interacting term, then a straightforward diagonalization is not an easy task. To overcome such situation, 
 we introduce the time-dependent rotation with an  angle $ \alpha(t) $
\begin{eqnarray}\label{rotation}
{\mathcal{R}_{\alpha}(t)=\exp[-i\alpha (t)\hat{L}_z]},
\qquad \alpha(t)=\frac{1}{2}\arctan\left( \frac{2J(t)}{\omega_{1}^{2}(t)-\omega_{2}^{2}(t)}\right)	
\end{eqnarray}
 in terms of the angular momentum $ {\hat{L}_z= \hat{x}_{1}\hat{p}_{2}-\hat{x}_{2}\hat{p}_{1}} $.
Consequently, 
the transformed Hamiltonian is given by 
\begin{eqnarray}\label{transham}
\tilde{H}(\hat{x}_{1},\hat{x}_{2})=  \mathcal{R}_{\alpha}(t) H(\hat{x}_{1},\hat{x}_{2}) \mathcal{R}_{\alpha}^{-1}(t)-i\mathcal{R}_{\alpha}(t)\partial_{t}\mathcal{R}_{\alpha}^{-1}(t)
\end{eqnarray}
and after some algebras, we obtain
\begin{eqnarray}\label{reham}
 {\tilde{H}(\hat{x}_{1},\hat{x}_{2})= \frac{\hat{p}_{1}^{2}}{2}+\frac{\hat{p}_{2}^{2}}{2}+\frac{1}{2}\Omega_{1}^{2}(t)\hat{x}_{1}^{2}+\frac{1}{2}\Omega_{2}^{2}(t)\hat{x}_{2}^{2}+\dot{\alpha}(t)(\hat{x}_{1}\hat{p}_{2}-\hat{x}_{2}\hat{p}_{1})}
\end{eqnarray}
where we have defined the frequencies $\Omega_{j}$ 
\begin{eqnarray}
\Omega_{1,2}^{2}(t)=\frac{1}{2}\left(\omega_{1}^{2}(t)+\omega_{2}^{2}(t)\pm
\sqrt{\left[\omega_{1}^{2}(t)-\omega_{2}^{2}(t)\right]^{2}+4 J^{2}(t)}\label{Omega}\right).
\end{eqnarray}
For the boundness  of Hamiltonian, the physical parameters point $ \mathcal{P}(\omega_{1},\omega_{2},J) $ should belongs to the  physical $ 3D$-space 
\begin{equation}\label{bound}
	\mathcal{E}_{B}=\left\lbrace \mathcal{P}/\ \ \omega_{1}^{2}\omega_{2}^{2}>J^{2}\right\rbrace
\end{equation}
It is clearly seen from
\eqref{reham}
 that the separation of variables is possible for  $ \dot{\alpha}(t)=0 $, 
 which is equivalent to have 
\begin{eqnarray}
{\label{condition} 
\tan (2\alpha(t))=\frac{2J(t)}{\omega_{1}^{2}(t)-\omega_{2}^{2}(t)}=\const}
\end{eqnarray}
which has been used also in different occasions, one may see  for instance \cite{R2, R17}.
Then 
the Hamiltonian 
is decoupled 
and canonically is equivalent to  the following time-dependent harmonic oscillators
\begin{eqnarray}
\tilde{H}(\hat{x}_{1},\hat{x}_{2})=\frac{\hat{p}_{1}^{2}}{2}+\frac{1}{2}\Omega_{1}^{2}(t)\hat{x}_{1}^{2}+\frac{\hat{p}_{2}^{2}}{2}+\frac{1}{2}\Omega_{2}^{2}(t)\hat{x}_{2}^{2}=\tilde{H}_{1}(\hat{x_{1}})+ \tilde{H}_{2}(\hat{x_{2}}) \label{ham1}
\end{eqnarray}
which can easily be solved to extract the solutions of  energy spectrum
and then solve different issues related to our system.

\subsection{Time-dependent Schr\"odinger equation } 

The commutativity $  \left[\tilde{H}_{1},\tilde{H}_{2} \right]=0 $ implies
that 
the solutions of time-dependent Schr\"{o}dinger equation have the forms
\begin{eqnarray}
\tilde\Psi(x_{1},x_{2};t)=\tilde\phi_{1}(x_{1};t)\otimes\tilde\phi_{2}(x_{2};t)
\end{eqnarray}
where each $ \tilde{\phi}(x_{j};t) $ satisfies
\begin{eqnarray}
\left( -\frac{1}{2}\partial_{x_{j}}^{2}+\frac{1}{2}\Omega_{j}^{2}(t)\hat{x}_{j}^{2}\right)\tilde{\phi_{j}} (x_{j};t)=-i\partial_{t}\tilde{\phi}_{j}(x_{j};t),\qquad j=1,2.
\end{eqnarray}
This was earlier studied in \cite{R5} an then the general solution of the Schr\"{o}dinger equation is a superposition of orthonormal  expanding modes $ \tilde \psi(x_{j};t)= \sum\limits_{n=0} \mathfrak{p}_{n_{j}}(t)\tilde{\phi}_{j}(x_{j},t) $, with $ \sum\limits_{n_{j}=0}|\mathfrak{p}_{n_{j} }(t)|^{2}=1 $. It follows that for a single mode  and $\Omega_{j}^{2}(0)>0$, we have
\begin{eqnarray}
\tilde{\phi_{j}}(x_{j};t)= \exp\left[-i\left( n_{j}+\frac{1}{2}\right) \int_{0}^{t}\varpi_{j}(\tau) d\tau \right] \chi_{n_{j}}\left( x_{j};t\right)
\end{eqnarray}
and  the orthogonal Hermite polynomials are
\begin{eqnarray}
\chi_{n_{j}}(x_{j};t)= \frac{1}{\sqrt{2^{n_{j}}n_{j}!}}\left( \frac{\varpi_{j}(t)}{\pi}\right)^{\frac{1}{4}} \exp\left[-\frac{1}{2} \varpi_{j}(t)x_{j}^{2}\right] \mathcal{H}_{n_{j}}\left(\sqrt{\varpi_{j}(t)} x_{j} \right)
\end{eqnarray}
where we have defined the scaling frequencies as $ \varpi_{j}(t)=\frac{\Omega_{j}(0)}{h_{j}^{2}(t)} $. 
The functions $ h_{j}(t) $ are  solutions of  Ermakov equations (dots stand for time
derivatives hereafter)
\begin{eqnarray}
\ddot{h}_{j}+\Omega_{j}^{2}(t)h_{j}=\frac{\Omega_{j}^{2}(0)}{h_{j}^{3}} \label{ermakov}
\end{eqnarray}
and satisfy the initial conditions $ h_{j}(0)=1, \dot{h}_{j}(0)=0 $.
It is worthy to note that the energy spectrum is time-independent  
\begin{eqnarray}
E_{n_{j}}= \left(n_{j}+\frac{1}{2} \right)\Omega_{j}(0) 
\end{eqnarray}
whereas the average of energy is time-dependent because we have
\begin{equation}\label{avH}
	\left\langle \tilde{H}(x_{j};t) \right\rangle_{n_{j}} =\frac{2n_{j}+1}{4\Omega_{j}(0)}\left(\dot{h}_{j}^{2}+\Omega_{j}^{2}(t)h_{j}^{2}+\frac{\Omega_{j}^{2}(0)}{h_{j}^{2}}\right).
\end{equation}
 Consequently the eigenfunctions 
 of   decoupled {Hamiltonian} are given by
\begin{eqnarray}\nonumber
\tilde{\Psi}_{n_1,n_2}(x_{1},x_{2};t)&=& \frac{1}{\sqrt{2^{n_{1}+n_{2}}n_{1}!n_{2}!}}\left( \frac{\varpi_{1}(t)\varpi_{2}(t)}{\pi^{2}}\right)^{\frac{1}{4}}\mathcal{H}_{n_{1}}\left(\sqrt{\varpi_{1}(t)} x_{1} \right)\mathcal{H}_{n_{2}}\left(\sqrt{\varpi_{2}(t)} x_{2} \right)\\
&& \times \exp\left[-i\left( n_{1}+\frac{1}{2}\right) \int_{0}^{t}\varpi_{1}(\tau) d\tau -i\left( n_{2}+\frac{1}{2}\right) \int_{0}^{t}\varpi_{2}(\tau) d\tau\right]\\
&& \times \nonumber  \exp\left[\frac{i}{2}\left(\frac{\dot{h}_{1}}{h_{1}}-i\varpi_{1}(t)\right) x_{1}^{2}+\frac{i}{2}\left( \frac{\dot{h}_{2}}{h_{2}}-i\varpi_{2}(t) \right) x_{2}^{2} \right].   
\end{eqnarray}
Now by performing the reciprocal rotation $ \mathcal{R}_{-\alpha}(t) $, we end up with the single mode solution of the Hamiltonian~(\ref{ham}), which is 
\begin{eqnarray}\nonumber
\Psi_{n_1,n_2}(x_{1},x_{2};t)&=&\mathcal{R}_{-\alpha}(t)\tilde{\psi}_{n_1,n_2}(x_{1},x_{2};t)\\
&=& \tilde{\Psi}_{n_1,n_2}\left(x_{1}\cos\alpha-x_{2}\sin\alpha,x_{1}\cos\alpha+x_{2}\sin\alpha;t  \right).
\end{eqnarray} 
In the forthcoming analysis, we only consider  the following vacuum solution
\begin{eqnarray}\nonumber
\Psi_{0,0}(x_{1},x_{2};t)&=& \left( \frac{\varpi_{1}(t)\varpi_{2}(t)}{\pi^{2}}\right)^{\frac{1}{4}}\  \exp\left[-\frac{i}{2} \int_{0}^{t}\varpi_{1}(\tau) d\tau -\frac{i}{2} \int_{0}^{t}\varpi_{2}(\tau) d\tau\right]\\ 
&&\times \exp\left[-\frac{1}{2} \mathcal{A}_{1}(t)x_{1}^{2}-\frac{1}{2}\mathcal{A}_{2}(t)x_{2}^{2}+\mathcal{A}_{12}(t)x_{1}x_{2}\right]\label{solution} 
\end{eqnarray}
where the involved time-dependent parameters read as 
\begin{eqnarray}
	\label{A11}
\mathcal{A}_{1}(t)&=& \varpi_{1}(t)\cos^{2}\alpha+\varpi_{2}(t)\sin^{2}\alpha-i\left(\frac{\dot{h}_{1}}{h_{1}} \cos^{2}\alpha+\frac{\dot{h}_{2}}{h_{2}} \sin^{2}\alpha \right) \\
\mathcal{A}_{2
}(t)&=& \varpi_{1}(t)\sin^{2}\alpha+\varpi_{2}(t)\cos^{2}\alpha-i\left(\frac{\dot{h}_{1}}{h_{1}} \sin^{2}\alpha+\frac{\dot{h}_{2}}{h_{2}} \cos^{2}\alpha\right) \label{A22}\\ 
\mathcal{A}_{12
}(t)&=& \sin\alpha\cos\alpha\left[\varpi_{1}(t)-\varpi_{2}(t) +i 
\left(\frac{\dot{h}_{2}}{h_{2}}-\frac{\dot{h}_{1}}{h_{1}} \right)\right]\label{A12}.
\end{eqnarray}
These results will be used to 
compute the number of occupation and discuss entanglement
trough the logarithmic negativity. 

\section{Classical stability and differential equations \label{Sec3}}

As we have seen above the vacuum state \eqref{solution} is strongly depending on 
the functions $ h_{1} $ and $ h_{2} $ solutions of Ermarkov equations \eqref{ermakov}. 
 Then it is of interest to discuss the classical stability and instability  related to Ermakov equation by   deriving the conditions of  classical stabilities. In the beginning, let us  perform  the following transformation on $ h_{1} $ and $ h_{2} $  \cite{korsch98}
\begin{eqnarray}
\mathbb{X}(t)&=& h_{1}(t)\ e^{i\int_{0}^{t}\varpi_{1}(s)ds}\\ \mathbb{Y}(t)&=&h_{2}(t)\ e^{i\int_{0}^{t}\varpi_{2}(s)ds}
\end{eqnarray}
to obtain the Hill system 
\begin{eqnarray}\label{eq1}
\ddot{\mathbb{X}}+\Omega_{1}^{2}(t)\mathbb{X}&=0\\
\ddot{\mathbb{Y}}+\Omega_{2}^{2}(t)\mathbb{Y}&=0 \label{eq2}
\end{eqnarray}
describing the two decoupled classical time-dependent  harmonic oscillators
of frequencies $ \Omega_{1}(t) $ and $ \Omega_{2}(t) $. Now it is clear that the stability of Hill system solutions leads to find that of Ermakov one
and therefore 
$ h_{j}(t) $ can be expressed as \cite{R9}
\begin{eqnarray}
h_{1}^{2}(t) &=&  x_{1}^{2}(t)+\Omega_{1}^{2}(0)
W^{-2}[x_{1},x_{2}] x_{2}^{2}(t)\\
 h_{2}^{2}(t) &=&  y_{1}^{2}(t)+\Omega_{2}^{2}(0)W^{-2}[y_{1},y_{2}] y_{2}^{2}(t).\label{Pinney}
\end{eqnarray}
such that 
 $ x_{j} $ and $ y_{j} $ are  independent solutions of (\ref{eq1}) and (\ref{eq2}), respectively, satisfying the {\color{red}initial }conditions  $ x_{1}(0)=y_{1}(0)=1 $ and $x_{2}(0)=y_{2}(0)=0$. Both Wronskian $ W[x_{1},x_{2}]=x_{1}\dot{x}_{2}-x_{2}\dot{x}_{1} $ and  $ W[y_{1},y_{2}]=y_{1}\dot{y}_{2}-y_{2}\dot{y}_{1} $ are constant.
 
\begin{figure}[H]
 	\centering
 	\includegraphics[width=6.26cm, height=8.5cm]{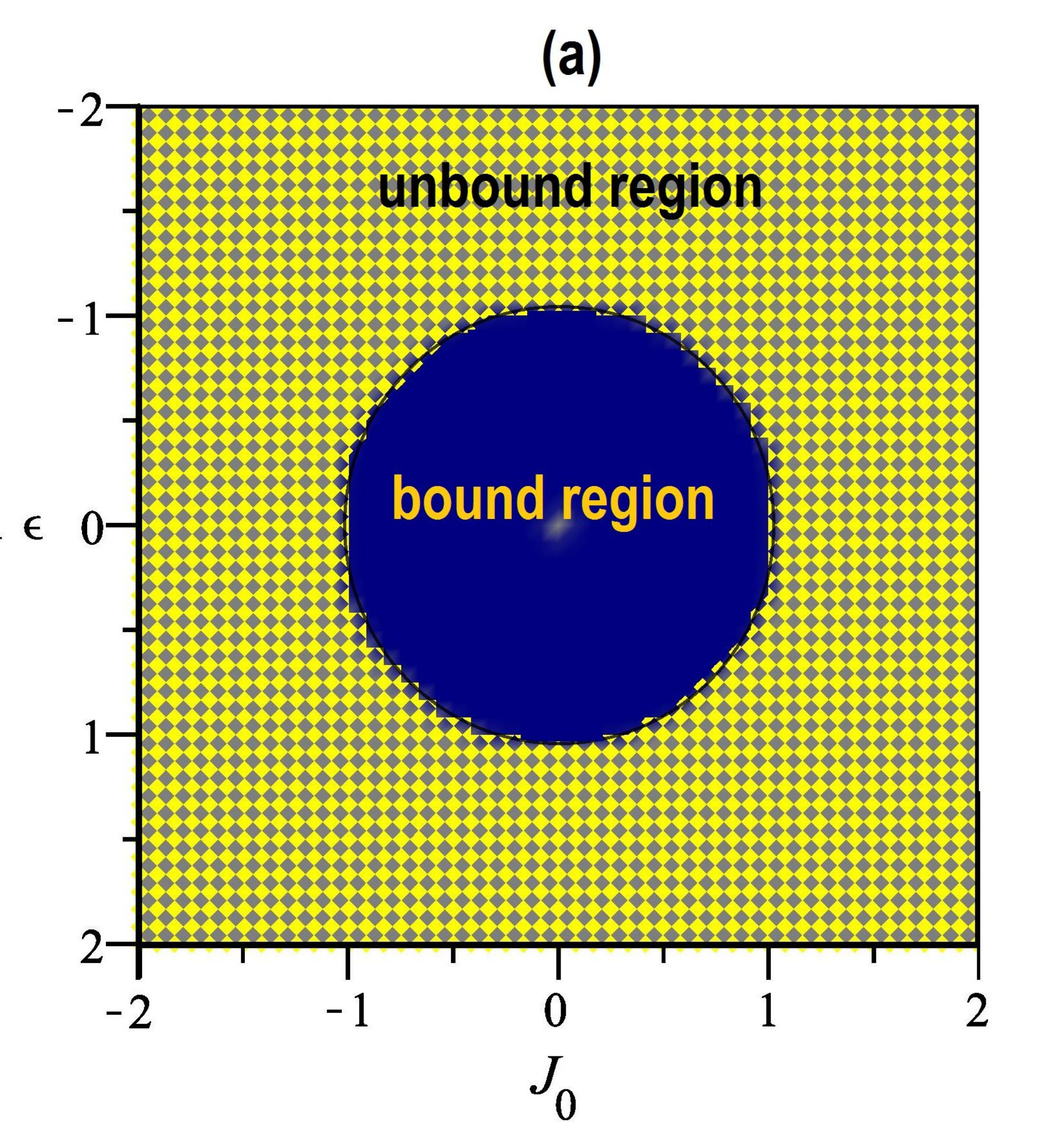}\ \ \ \ \ \ \
 	\includegraphics[width=6.5cm, height=8.9cm]{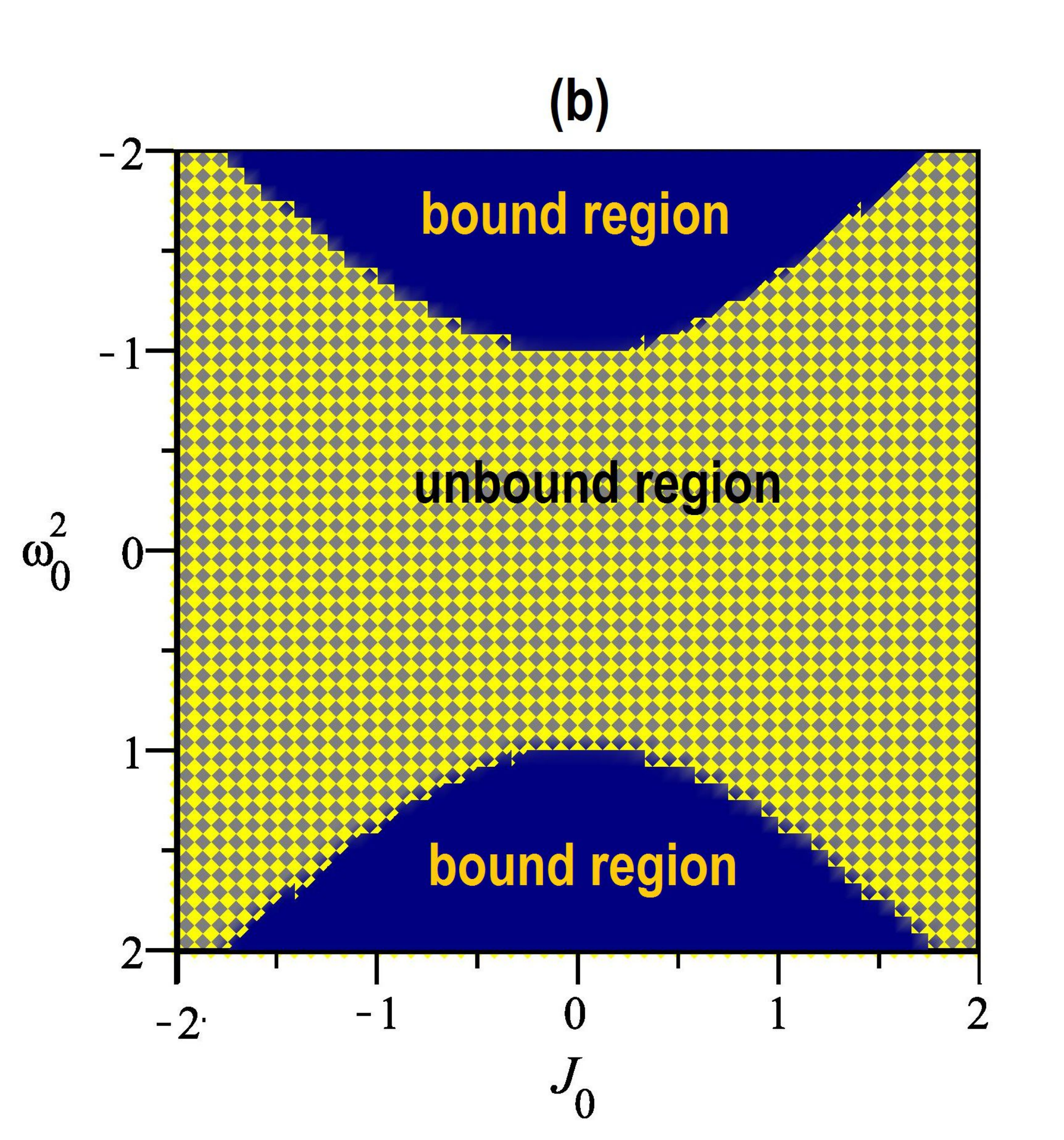}
 	\captionof{figure}{\sf (color online) The boundness maps in two space configurations. (a): Configuration $ (\epsilon,J_{0})$ for $\omega_{0}^{2}=1$,  as predicted the physical points that bound the Hamiltonian (\ref{ham}) form an open disc of radius $ R=\omega_{0}^{2}=1$. (b): Configuration  $  (\omega_{0}^{2},J_{0})$ for $ \epsilon=1 $. \label{bound1}}
 \end{figure}

 To proceed further, we require some conditions on frequencies and
 coupling parameter \cite{R10,R15}. Indeed, let us modulate them as
\begin{eqnarray}\label{modu}
\omega_{1,2}^{2}(t)=\omega_{0}^{2}\pm\epsilon \Theta(t) 
,\qquad J(t)= J_{0} \Theta(t)
\end{eqnarray}
where $ \omega_{0} $,  $ J_{0} $, and $\epsilon$ are  the parametric  frequency,   coupling amplitude and     quench amplitude, respectively. The
involved  $\pi- $periodic quencher $ \Theta(t)$ is defined by
\begin{eqnarray}
\Theta(t) =\left\lbrace \begin{array}{ll}
+1& 0\leq t<\frac{\pi}{2}\\
-1& \frac{\pi}{2}\leq t< \pi
\end{array}
\right.
\end{eqnarray}
and then the frequencies \eqref{Omega} reduce to the following
\begin{eqnarray}
\Omega_{1,2}^{2}(t)=\omega_{0}^{2}\pm\Theta(t) \sqrt{\epsilon^{2}+J_{0}^{2}}.
\end{eqnarray}
It is worthy to note that with the modulation \eqref{modu}, the Hill system (\ref{eq1}-\ref{eq2})  reduces to the Meissner equations \cite{R15}. 
Now, we discuss the boundness of Hamiltonian because the solutions presented in (\ref{solution}) are only valid  for $ \Omega_{j}(0)>0$, with $ j=1,2 $  \cite{R11}. As a result  of \eqref{modu} the boundness condition (\ref{bound}) becomes $\omega_{0}^{4}-\epsilon^{2}>J_{0}^{2}$, which is equivalent to an open disc of center $ (J_{0}^{2} = 0, \epsilon^{2} = 0) $ and radius $R = \omega_{0}^{2}$. In {\color{blue}Figure} \ref{bound1},
we give the physical maps for boundness of Hamiltonian in two different configurations $ (\epsilon,J_{0}) $ and $ (\omega_{0}^{2},J_{0}) $. The  maps show that the unbound regions are
very large than bound ones, which limits our choices. Note that, the edges of boundness present great importance for example in generating important entanglement and  leading to inverse engineering of time-dependent coupled harmonic oscillators \cite{R12}.

\begin{figure}[H] 
	\centering
	\includegraphics[width=7cm, height=
	7cm]{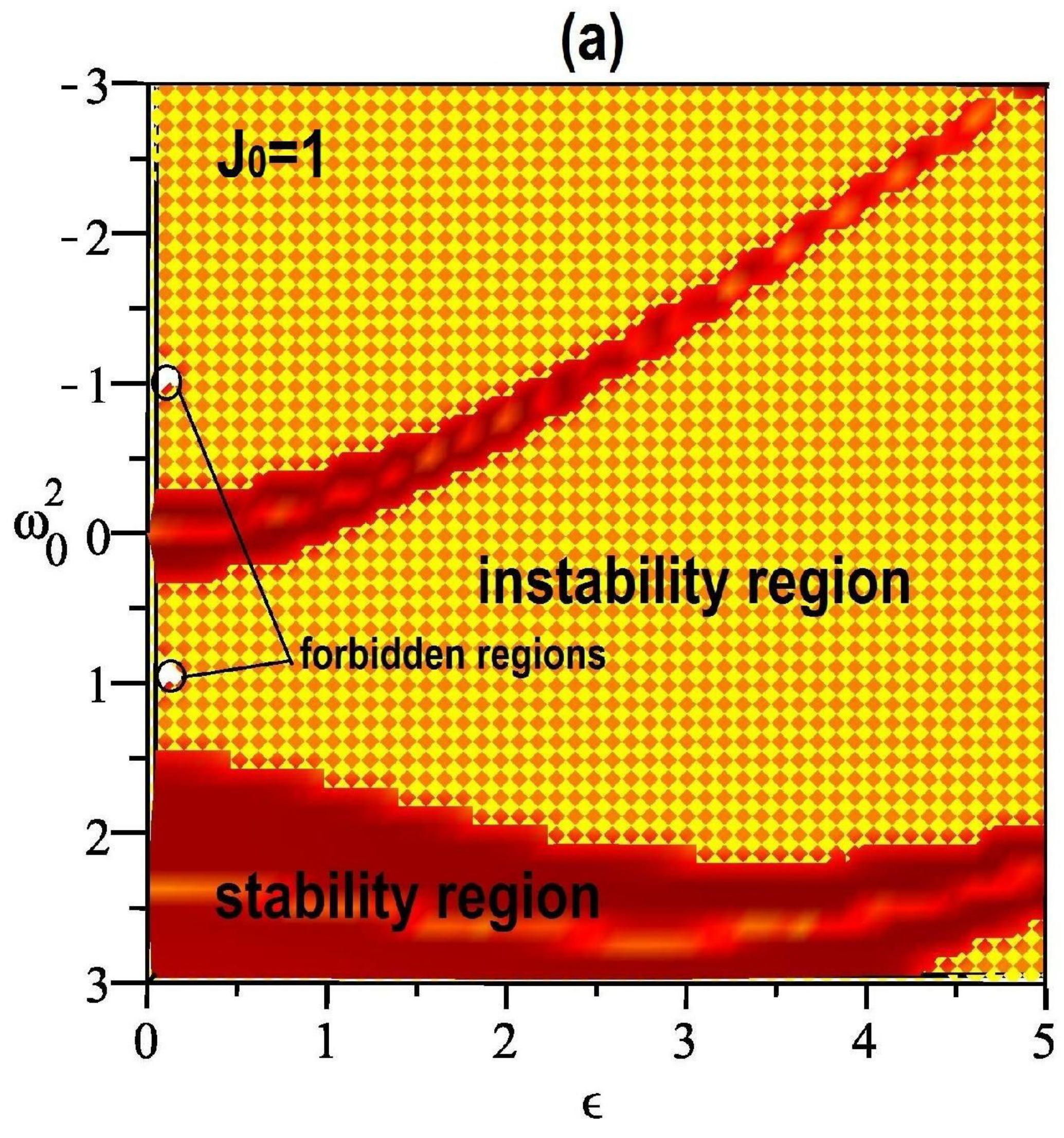}\ \ \ \ \ \ \
	\includegraphics[width=7cm, height=
	7.3cm]{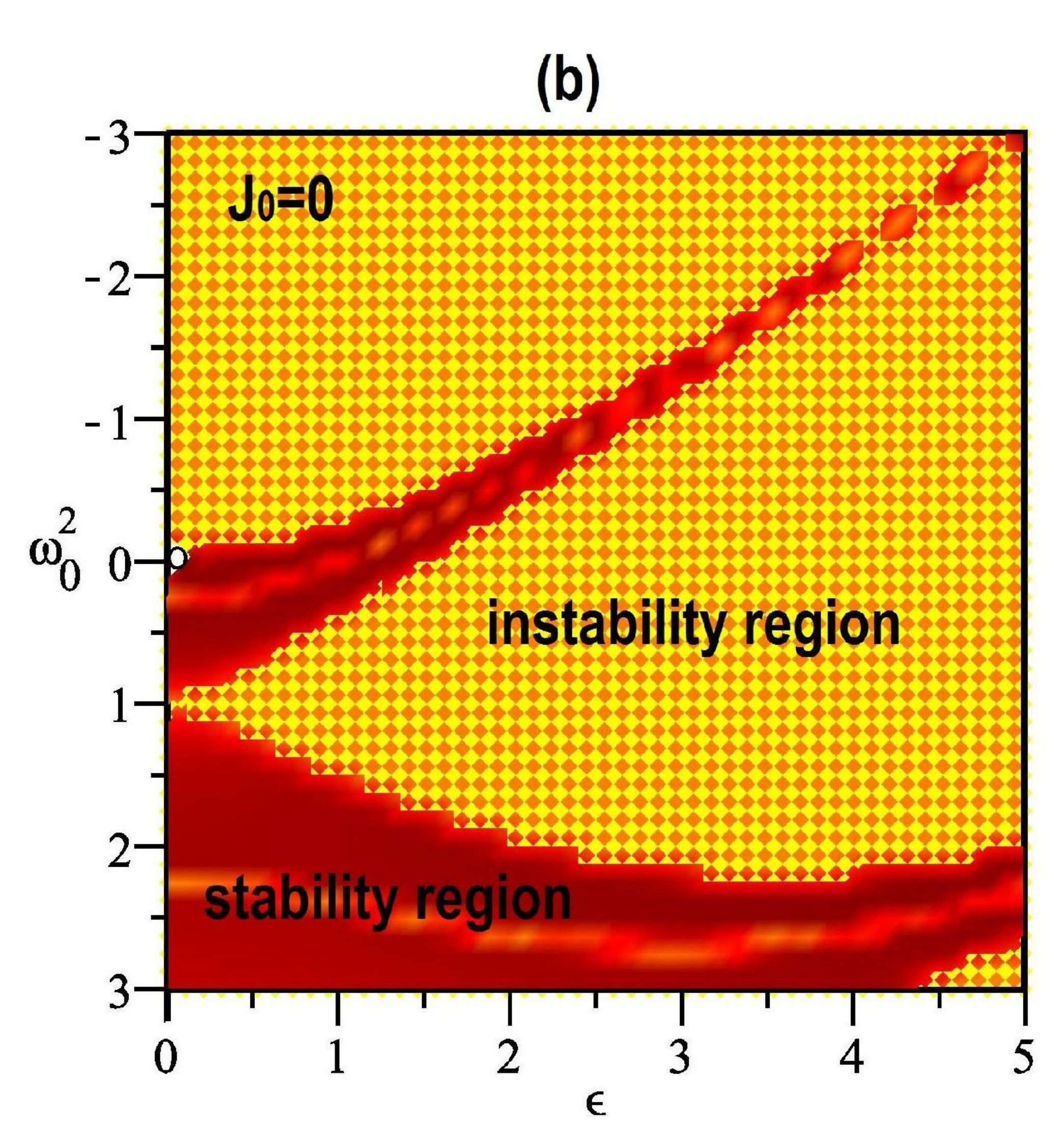}
	\captionof{figure}{\sf (color online) The stability diagram of Meissner system (\ref{eq1}-\ref{eq2}) with modulation \eqref{modu} in the space configuration $(\epsilon,\omega_{0})$ for two cases (a): $ J_{0}=0 $ and (b): $ J_{0}=1 $. The yellow area stands for instable region $ (\mathcal{S}=0) $,  red area  for stable region $ ( \mathcal{S}>0) $, white dots for forbidden parameters ($\Omega_{2}(0)=0$).\label{stable} 	  }
\end{figure}

To study the classical instability of (\ref{eq1}-\ref{eq2}), we will use the discrete transition matrix formalism or Floquet exponents 
technique \cite{R10}. Then, after some algebra we show that  of the stability condition of  two oscillators can be written as
\begin{eqnarray}
	\mathcal{S}:=\max\left[1-\Lambda(\Omega_{1},\Omega_{2}),0 \right]>0
\end{eqnarray}
where $\Lambda$ is a dimensionless parameter
\begin{eqnarray}
	\Lambda(\Omega_{1},\Omega_{2})=\left| \cos\left( \frac{\Omega_{1}\pi}{2}\right) \cos\left( \frac{\Omega_{2}\pi}{2}\right)-\frac{1}{2}\left(\frac{\Omega_{1}}{\Omega_{2}}+\frac{\Omega_{2}}{\Omega_{1}}\right)\sin\left( \frac{\Omega_{1}\pi}{2}\right)\sin\left( \frac{\Omega_{2}\pi}{2}\right)\right| \label{stabb}.
\end{eqnarray}
In {\color{blue} Figure} \ref{stable}, we numerically show the stability diagram in the  physical configuration $ (\epsilon,\omega^{2}_{0}) $. It is clearly seen that  the stability diagram is very sensitive to the physical  parameters because a small change  produces important configuration variation. We notice  that the instability region increases as long as the coupling parameter increases. This in fact tells us that  why one has to investigate the effect of instabilities on the quantum features.

\section{Entanglement and photon excitation's \label{sec4}}
\subsection{Entanglement and dynamics effects} 

According to Peres-Horodecki criterion, 
\cite{R1,R6}, 
the necessary and sufficient condition for the separability of two  Gaussian mode states is the positivity of the partially transposed state. 
Since the vacuum state \eqref{solution} is pure and symmetric, then 
separability can be realized by switching  \eqref{A12} to zero, namely 
having the condition
\begin{eqnarray}
\mathcal{A}_{12}(t)=0\label{separability}
\end{eqnarray}
which
 is necessary and sufficient for separability and obviously it is achieved in two subordinate cases, 
 (i): 
 for $ J_{0}=0 $   the dynamics can not generate entanglement and the oscillators still separable during the dynamics, (ii): 
 for the Wronskian $ W[h_{1}(t),h_{2}(t)]=0 $ and $\varpi_{1}(t)=\varpi_{2}(t)$ where the geometrical meaning of $ W $ is substraction of  the rectangular phase space areas $ S_{1}= h_{1}\dot{h}_{2} $ and $ S_{2}= h_{2}\dot{h}_{1}$ \cite{R12}. The last case indicates that the  dynamics can extinct entanglement and  then to avoid  its extinction, the 
engineering {of initial} and final normal frequencies $\Omega_{j}(t_{f}), \, \Omega_{j}(0)$ $ (j=1,2) $ deserves a suitable tuning.

Regarding our case,  the vacuum state is completely described by the marginal purities $ \mu_{j} $, $ (j=1,2) $ given by 
\begin{eqnarray}
\mu_{1}(t)=\mu_{2}(t):=\mu(t)=\left( \frac{\varpi_{1}(t)\varpi_{2}(t))}{\varpi_{1}(t)\varpi_{2}(t)+|\mathcal{A}_{12}(t)|^{2}}\right)^{\frac{1}{2}}.
\end{eqnarray}
Since our state is  pure and  Gaussian, then all  quantum correlation can be derived from the second moment of it, 
that is  the covariance matrix (CM) $\mathcal{V}(t)$. Such CM  can be transformed via a local symplectic transformation $ S=S_{1}\oplus S_{2} $ to a particular form called standard form $\mathcal{V}_{sf}(t)$ \cite{R1}
\begin{eqnarray}
\mathcal{V}_{sf}(t)&=&\left( \begin{array}{cccc}
	\mu^{-1} & 0 & \sqrt{\mu^{-2}-1} & 0 \\ 
0 & \mu^{-1} & 0 & -\sqrt{\mu^{-2}-1} \\ 
\sqrt{\mu^{-2}-1} & 0 & \mu^{-1} & 0 \\ 
0 & -\sqrt{\mu^{-2}-1} & 0 & \mu^{-1}
\end{array}\right)=\left( \begin{array}{cc}
A & C \\ 
C & A
\end{array} \right).
\end{eqnarray}
By performing the {partial transposition (PT)} prescription, $ \det(A)\longrightarrow \det(A) $ and $ \det(C)\longrightarrow-\det(C) $, then 
the minimal symplectic eigenvalue of the PT   covariance matrix $\tilde{\mathcal{V}}$ is
\begin{eqnarray}
\lambda_{min}^{2}(t)= \frac{1}{2}\left(\Delta(\tilde{\mathcal{V}})-\sqrt{\Delta^{2}(\tilde{\mathcal{V}})-4} \right)
\end{eqnarray} 
and after evaluation, we find
\begin{eqnarray}
\lambda_{min}^{2}(t)=\left(1+ \frac{|\mathcal{A}_{12}(t)|^{2}}{\varpi_{1}(t)\varpi_{2}(t)}\right)
\left( 1-\sqrt{1-\mu^{2}(t)}\right) -1
\end{eqnarray} 
where the symplectic invariant is $ \Delta(\tilde{\mathcal{V}}) =2(\det (A)-\det(C))$. Consequently, the logarithmic negativity is given by
\begin{eqnarray}
E_{\mathcal{N}}(t)= \max\left(0,-\log\left[\lambda_{min}(t)\right] \right) 
\end{eqnarray} 
which is monotonically increasing with $|\mathcal{A}_{12}|^{2}$.   From its expression (\ref{A12}) it appears that the main contributions of the time-dependent Hamiltonian is the emergence of an imaginary part, {that is $
\left(\frac{\dot{h}_{2}}{h_{2}}-\frac{\dot{h}_{1}}{h_{1}} \right) $} and  initial normal mode scaling, i.e. $\Omega_{j}(0)\longrightarrow \varpi_{j}(t)= \frac{\Omega_{j}(0)}{h_{j}^{2}(t)}$.

\subsection{Photon excitation's}

Using  of  the phase space prescription \cite{R13}, we will analyze photon excitation's by computing the average of  photon numbers $\langle a_{j}^{+}a_{j} \rangle$ in the vacuum state. The case of two resonant time-independent coupled oscillators was analyzed in \cite{R7} where 
the creation $ a^{+}_{j,0} $ and annihilation $ a_{j,0} $ operators are simply mapped as
\begin{eqnarray}
(a^{+}_{j,0})^{+}=a_{j,0}=\sqrt{\frac{\omega_{j}}{2}}x_{j}+\frac{i}{\sqrt{2\omega_{j}}}p_{j}.
\end{eqnarray}
However for  time-dependent Hamiltonian,  the situation is not obvious
because the realization can be done as follows \cite{R14,R8}
\begin{eqnarray}
(a^{+}_{j})^{+}=a_{j}=e^{i\int_{0}^{s}\eta_{j}(s)ds} \frac{1}{\sqrt{2\eta_{j}}}\left[\eta_{j}\left( 1-i\frac{\dot{\nu_{j}}}{\nu_{j}\eta_{j}}\right) x_{j}+ip_{j}\right]
\end{eqnarray} 
where   $\left[a_{j},a^{+}_{j}  \right]=\mathbb{I}$, $ \eta_{j}(t)=\frac{\omega_{j}(0)}{\nu_{j}^{2}} $ and  the  functions $ \nu_{j} $ satisfy the Ermakov equations 
\begin{eqnarray}
\ddot{\nu_{j}}+\omega_{j}^{2}(t)\nu_{j}=\frac{\omega_{j}^{2}(0)}{\nu_{j}^{3}},
\qquad  j=1,2.
\end{eqnarray}
After a straightforward algebra, one can compute the average of photon number operators $N_{j}= a_{j}^{+}a_{j}$ to end up with
\begin{eqnarray}
\left\langle N_{j}\right\rangle(t) = \frac{1}{2\omega_{j}(0)}\left(  \frac{\omega_{j}^2(0)}{\nu_{j}^2}+\dot{\nu}_{j}^{2}\right) \langle\hat{x}_{j}^{2}\rangle_{t}+\frac{\nu_{j}^{2}}{2\omega_{j}(0)}\langle\hat{p}_{j}^{2}\rangle_{t}-\frac{\nu_{j}\dot{\nu}_{j}}{\omega_{j}(0)}\langle\hat{x}_{j}\hat{p_{j}}\rangle_{t}-\frac{1}{2} \label{photon}
\end{eqnarray} 
and  different averages are explicitly given by 
\begin{eqnarray}
&&\langle\hat{x}_{j}^{2}\rangle_{t}= \frac{\varpi_{j}(t)\sin^{2}\alpha+\varpi_{k}(t)\cos^{2}\alpha}{2\varpi_{j}(t)\varpi_{k}(t)}\label{avx}\\
&&\langle\hat{p}_{j}^{2}\rangle_{t}=\frac{1}{2}\left[\varpi_{j}(t)\cos^{2}\alpha+\varpi_{k}(t)\sin^{2}\alpha  +\frac{1}{\varpi_{j}(t)}\left(\frac{\dot{h}_{j}}{h_{j}}\right)^{2}\cos^{2}\alpha+\frac{1}{\varpi_{k}(t)}\left(\frac{\dot{h}_{k}}{h_{k}}\right)^{2}\sin^{2}\alpha \right] \label{avp}\\
&& \langle\hat{x}_{j}\hat{p}_{j}\rangle_{t} =- \frac{\varpi_{k}(t)\cos^{2}\alpha\ \frac{\dot{h}_{j}}{h_{j}} +\varpi_{j}(t)\sin^{2}\alpha \frac{\dot{h}_{k}}{h_{k}}}{2\varpi_{j}(t)\varpi_{k}(t)} 
\end{eqnarray}
where $ k=1,2 $ and $ k\neq j $.

We emphasis that 
the time-dependent Hamiltonian (\ref{ham}) generates important effects such that the {scalings} $ \eta_{j}(t)=\frac{\omega_{j}(0)}{\nu_{j}^{2}} $ and 
$ \varpi_{j}(t)= \frac{\Omega_{j}(0)}{h_{j}^{2}(t)}$, the shiftings in positions \eqref{avx} and momenta \eqref{avp}, which are due to
the dilatation functions {$ \nu_{j}(t)$ and $ h_{j}(t)$}. In addition, the existence of the term $\langle\hat{x}_{j}\hat{p}_{j}\rangle_{t}$ 
is  purely a consequence of  time-dependence, which of course  
disappears by setting time to zero, namely having constant parameters. {Note that}, for time-independent Hamiltonian, we have $\nu_{j}=h_{j}=1$ and $\dot{\nu}_{j}=\dot{h}_{j}=0$, then the virtual photon excitation's become 
\begin{eqnarray}
\left\langle N_{1}\right\rangle &=&  \frac{1}{4}\cos^{2}\alpha\left( \frac{\Omega_{1}}{\omega_{1}}+\frac{\omega_{1}}{\Omega_{1}}\right)+\frac{1}{4}\sin^{2}\alpha\left( \frac{\Omega_{2}}{\omega_{1}}+\frac{\omega_{1}}{\Omega_{2}}\right)-\frac{1}{2}\\
\left\langle N_{2}\right\rangle &=&  \frac{1}{4}\sin^{2}\alpha\left( \frac{\Omega_{2}}{\omega_{2}}+\frac{\omega_{2}}{\Omega_{2}}\right)+\frac{1}{4}\cos^{2}\alpha\left( \frac{\Omega_{1}}{\omega_{2}}+\frac{\omega_{2}}{\Omega_{1}}\right)-\frac{1}{2}.
\end{eqnarray}
In the case of resonant oscillators, i.e. $\omega_{1}=\omega_{2}=\omega_{r}$, the rotation angle reduces to $ \alpha=\frac{\pi}{4} $. 
Furthermore, by setting $r_{a,b}=\frac{1}{2}\ln\left(\frac{\Omega_{1,2}}{\omega_{r}}\right) $, we retain the results derived in \cite{R7}
\begin{eqnarray}
\left\langle N_{1}\right\rangle =\left\langle N_{2}\right\rangle= \frac{1}{2}\left(\sinh^{2}r_{a}+\sinh^{2}r_{b}\right).
\end{eqnarray}
We mention that the study in \cite{R7} was confined  on the resonant case because 
it was not easy 
to disentangle the rotation operator (\ref{rotation}) in the frame of creation $a_{j}^{+} $ and annihilation $ a_{j} $ representation. Now it becomes clear that from our analysis how the  phase space picture can be used to overcome such situation. 
In addition, it is interesting to note that   for $\omega_{1}\neq\omega_{2}$ and $ J_{0}=0 $, the vacuum state \eqref{solution} contains excitation's, i.e. $ \left\langle N_{1}\right\rangle =\left\langle N_{2}\right\rangle\neq 0$, even if the oscillators are decoupled, such phenomena does not exist in the frame of resonant oscillators.  

\section{Results and discussions \label{sec5}}

Before  numerically presenting and discussing the main results derived so far, it is convenient for our task to
define dimensionless parameters  
\begin{eqnarray}
t\longrightarrow t\Omega_{0}, \qquad \omega_{j}\longrightarrow\frac{\Omega_{j}}{\Omega_{0}}, \qquad \omega_{j}\longrightarrow\frac{\omega_{j}}{\Omega_{0}},\qquad \epsilon \longrightarrow \frac{\epsilon}{\Omega^{2}_{0}},\qquad J_{0}\longrightarrow \frac{J_{0}}{\Omega^{2}_{0}}
\end{eqnarray} 
where $\Omega_{0}$ is an arbitrary   frequency. 
For a numerical study of classical instabilities effects on generation of photon excitation's and hence  entanglement between oscillators, we start  by giving the dilatation  functions
$ h_{j} $ and $ \nu_{j} $. Indeed, by using  (\ref{Pinney}) one can solve (\ref{ermakov}) to obtain  the solutions in  first period $ [0, \pi] $, which are 
\begin{eqnarray}
h_{1}^{2}(t)&=& \left\lbrace \begin{array}{ll}
1, & \ \ \ \ 0\leq t<\frac{\pi}{2}\\
\frac{\Omega_{2}^{2}-\Omega^{2}_{1}}{2\Omega_{2}^{2}}\cos\left(2\Omega_{2}(t-\frac{\pi}{2})\right)+\frac{\Omega_{1}^{2}+\Omega^{2}_{2}}{2\Omega_{2}^{2}}, &\ \ \ \ \frac{\pi}{2}\leq t< \pi
\end{array}
\right.\label{erm}\\
h_{2}^{2}(t)&=& \left\lbrace
\begin{array}{ll}
1,& \ \ \ \ 0\leq t<\frac{\pi}{2}\\
\frac{\Omega_{1}^{2}-\Omega^{2}_{2}}{2\Omega_{1}^{2}}\cos\left(2\Omega_{1}(t-\frac{\pi}{2})\right)+\frac{\Omega_{1}^{2}+\Omega^{2}_{2}}{2\Omega_{1}^{2}}, &\ \ \ \ \frac{\pi}{2}\leq t< \pi.
\end{array}
\right.\label{Erm}
\end{eqnarray}
Note that
the solutions of $\nu_{j}$ will be   immediately obtained from (\ref{erm}-\ref{Erm})  only under the substitution $\sqrt{\epsilon^{2}+J_{0}^{2}}\lga$  $\epsilon$. 
Now, 
we notice that 
the freezing dynamics will be occurred in  odd half periods while dynamics evolution in even ones,  which is due to the  periodic quench, and the continuity of solutions $ h_{j} $ and $ \nu_{j} $. 

\begin{figure}[H]
	\centering
	\includegraphics[width=7cm, height=7cm]{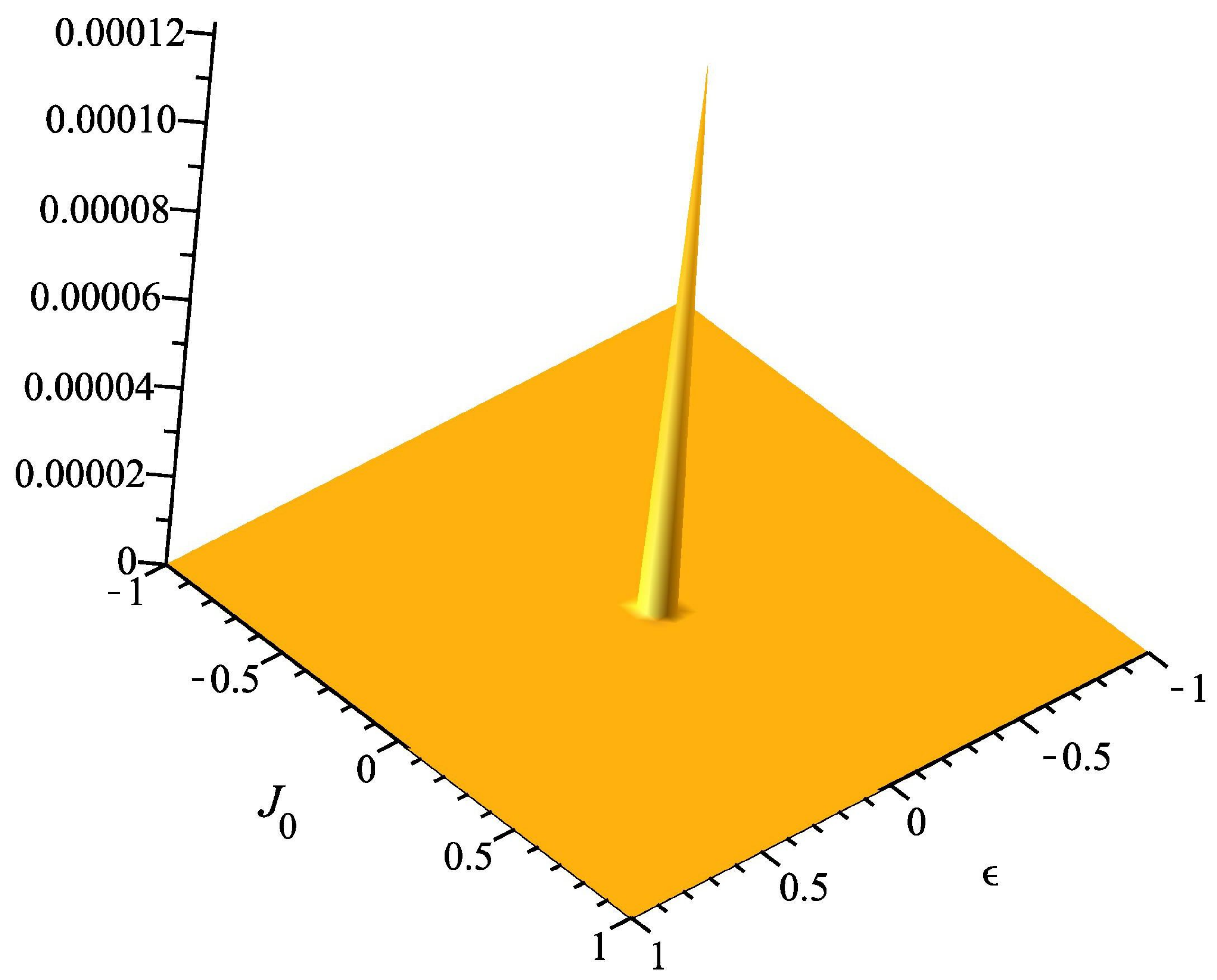}
	\includegraphics[width=7cm, height=7cm]{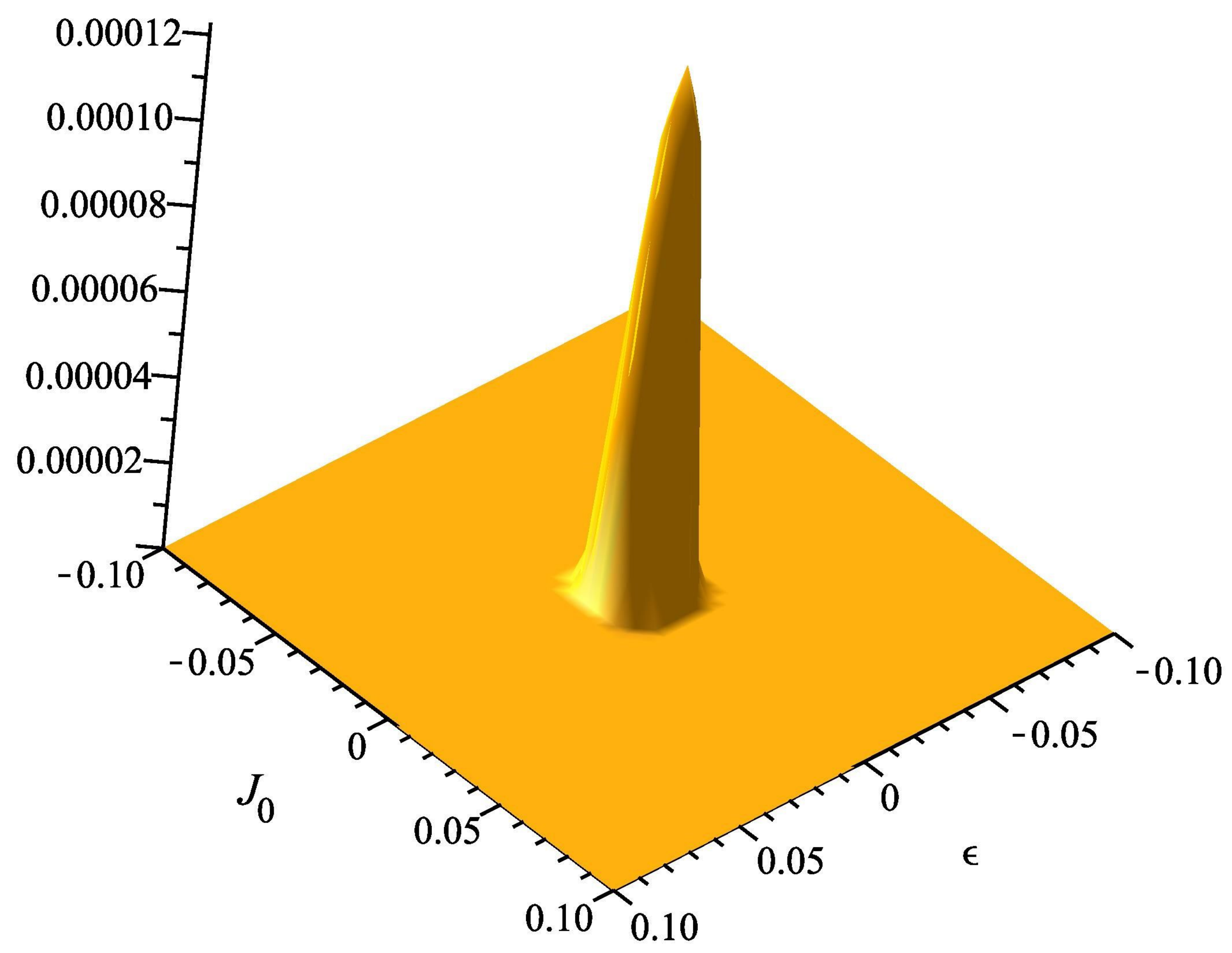}
	\includegraphics[width=4cm, height=4cm]{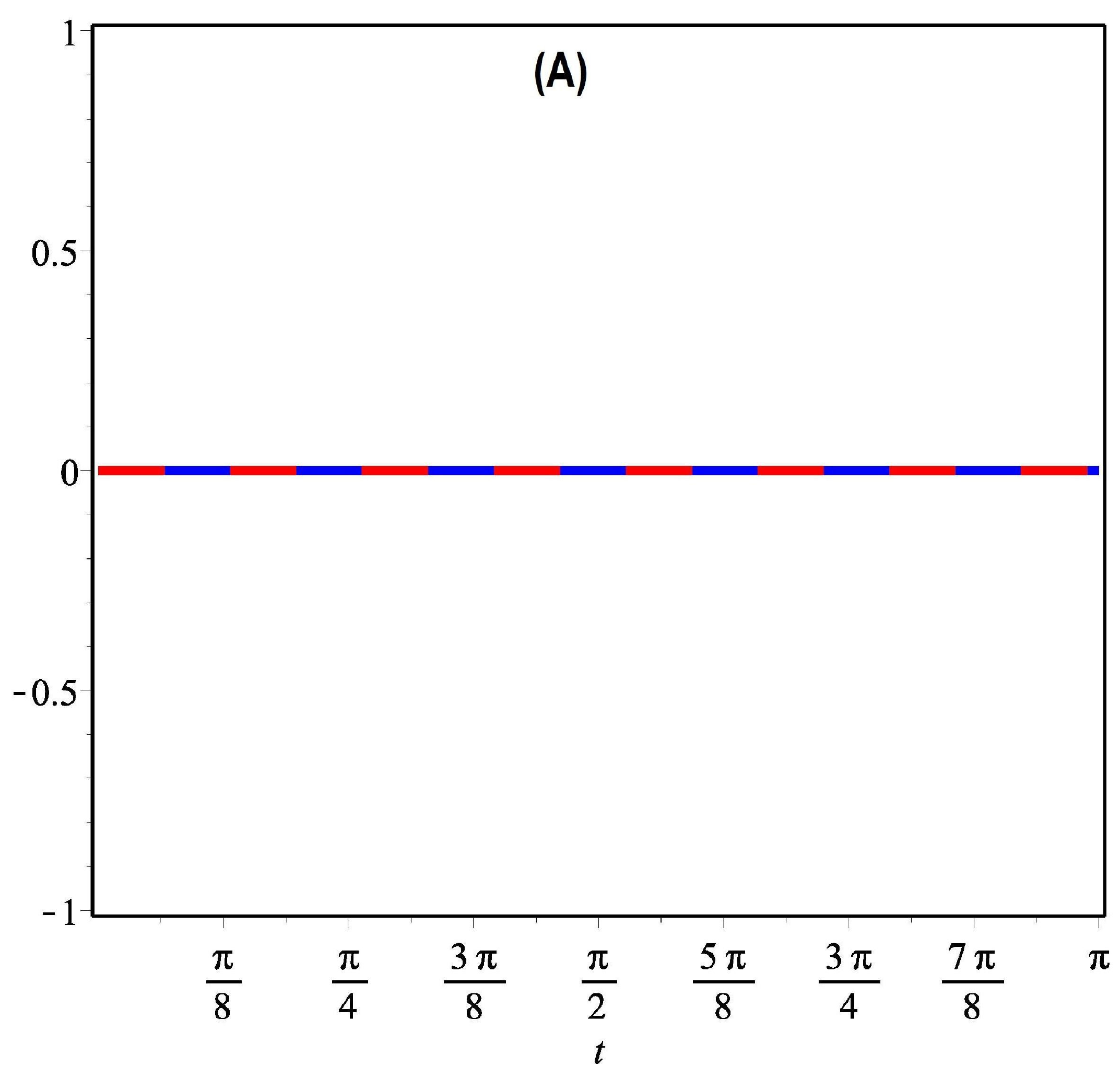}
	\includegraphics[width=4cm, height=4cm]{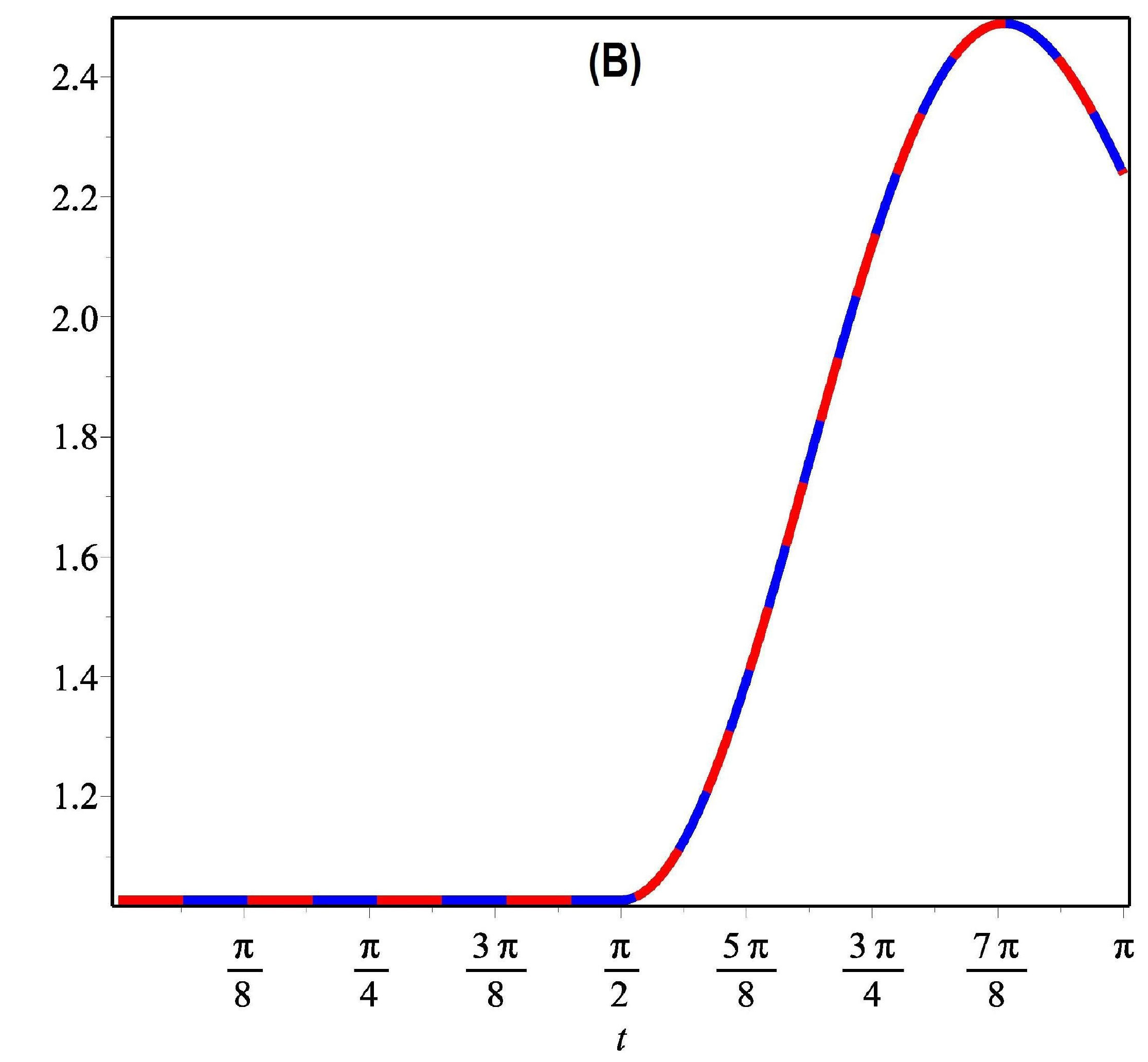}
	\includegraphics[width=4cm, height=4cm]{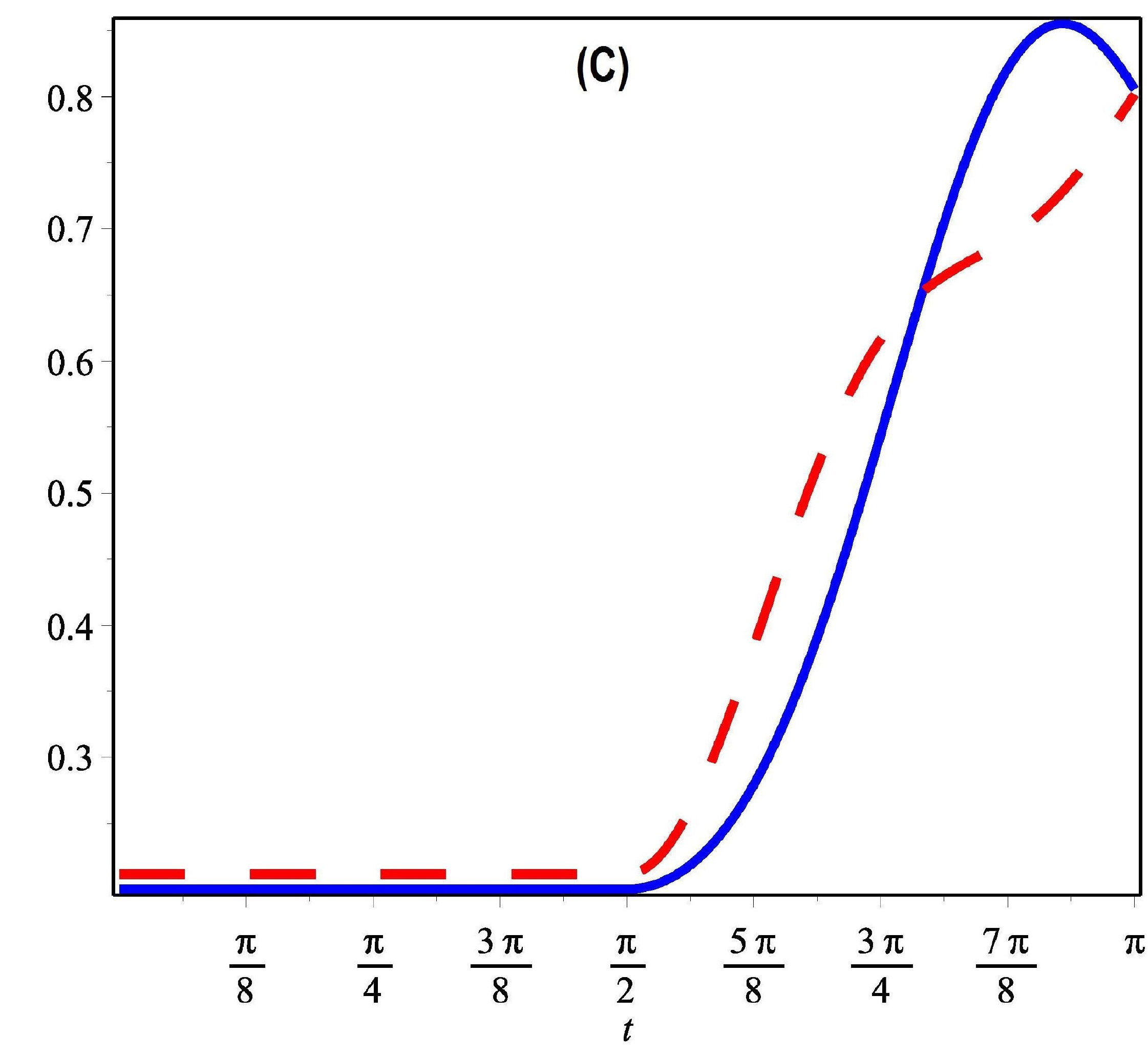}
	\includegraphics[width=4cm, height=4cm]{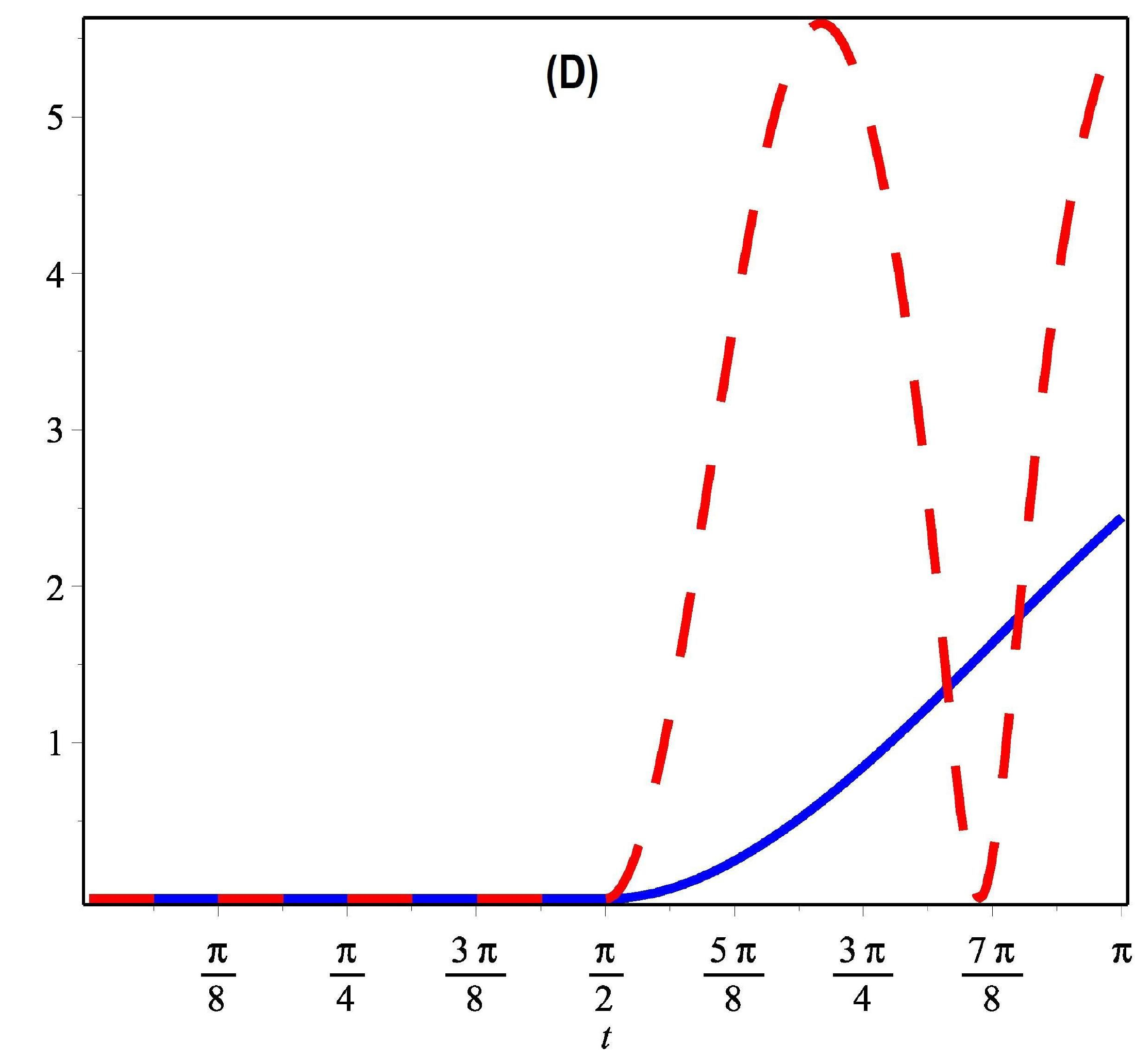}
	\captionof{figure}{\sf (color online)  The golden surface indicates instability region ($\mathcal{S}=0$) and the stability area ($\mathcal{S}>0$) for the frequency $\omega_{0}^{2}=1.01$. Panels A, B, C and D present the virtual photon excitation's  $ \langle N_{1}\rangle $ (blue line)   $ \langle N_{2}\rangle $ (red dashed) versus the time $t$. If we use the notation $ M(\epsilon,J_{0}) $, it turns out that  the physical points are   $ A(0,0) $, $ B(0,0.99) $, $ C(0.1,0.9) $,  and $ D(0.9,0.1) $. \label{photonstablity} 	  }
\end{figure}

 In {\color{blue} Figure} \ref{photonstablity}, we remark that vacuum state does not contain virtual excitation's when the classical analog of quantum oscillators are stable and vice verse. Another point that deserves attention is the emergence of excitation's even if the oscillators are decoupled $ (J_{0}=0)$ and beyond resonance $\omega_{1}\neq \omega_{2}$ ($ \epsilon\neq 0 $), amazingly the classical oscillators are unstable. Such phenomena  is not observed for resonant oscillators \cite{R7}.  Generally, the virtual photons are originated from counter-rotating (CR) terms ($ a_{1}^{+}a_{2}^{+} $ and $ a_{2}a_{1} $)  appearing  in the Hamiltonian. However, these   terms disappear when the coupling is switched-off and the non-resonance oscillations  has no relation with CR. Indeed, the resonance affects the   potential energy operator, $V(J_{0}=0)=\frac{1}{2}\omega_{1}^{2}(t)\hat{x}_{1}^{2}+\frac{1}{2}\omega_{2}^{2}(t)\hat{x}_{2}^{2}$, and this does not mix the quadratures $ \hat{x}_{j} $  and $ \hat{p}_{k}  $, then  CR terms does not appear. This phenomena can be seen as follows, when the classical oscillators are stable then
  the virtual excitation's will be suppressed, but for unstable case it
  will be created. Note that  $ \epsilon $ and $ J_{0} $ have the same dimension but  different effects on excitation's generation. In order to compare their effects on  excitation's amount and hierarchy, we also plot in {\color{blue} Figure} \ref{photonstablity}, the dynamics of excitation's  $ \langle N_{1}\rangle $ and $ \langle N_{2}\rangle $. The numerical results  show that the increasing of  $\epsilon$  increases the amount of excitation's as well as the  hierarchy $\langle N_{1}\rangle\neq\langle N_{2}\rangle$ and changes the topological behavior in the simulate time (linear/sinusoidal). Now, if  the values of $\epsilon$ and $ J_{0} $ $ (C\leftrightarrow D) $ are interchanged, we  obtain different aspects, which are due to  contributions of $\epsilon$ and  $ J_{0} $ in the frequencies $\omega_{j}$ and $\Omega_{j}$.
 
 In this way, we can affirm that virtual photons are correlated with the instability of classical counterpart's solutions. In our best of knowledge, this phenomena is the first time that it has been studied, and this will lead to control virtual excitation's in quantum systems only by  engineering  their classical counterparts, where the financial requirements are not enormous. Also it  will leads  to understand the nature of the connection between quantum systems and theirs classical counterparts.\\

\begin{figure}[H]
	\centering
	\includegraphics[width=6cm, height=6cm]{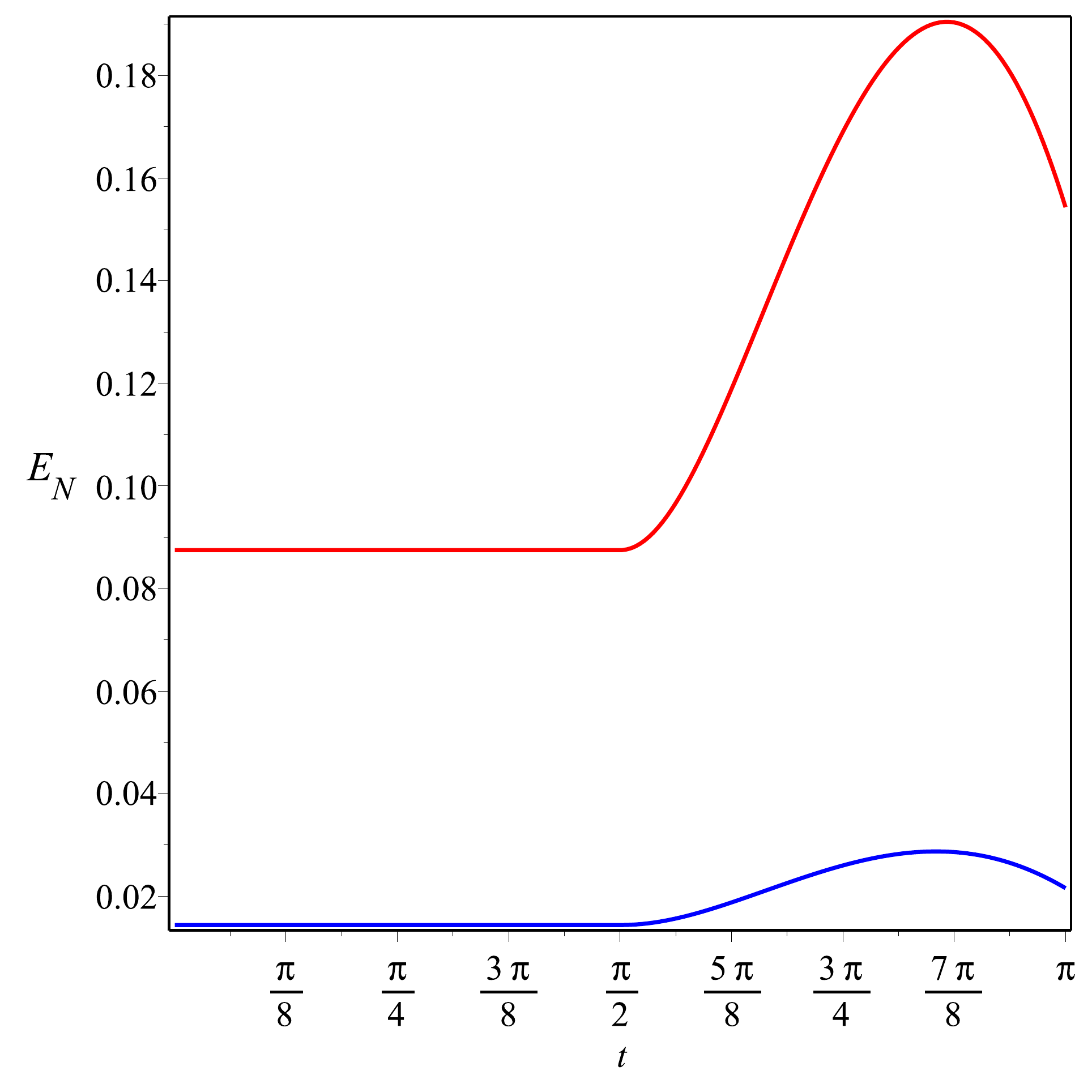}\ \ \ \ \
	\includegraphics[width=6cm, height=6cm]{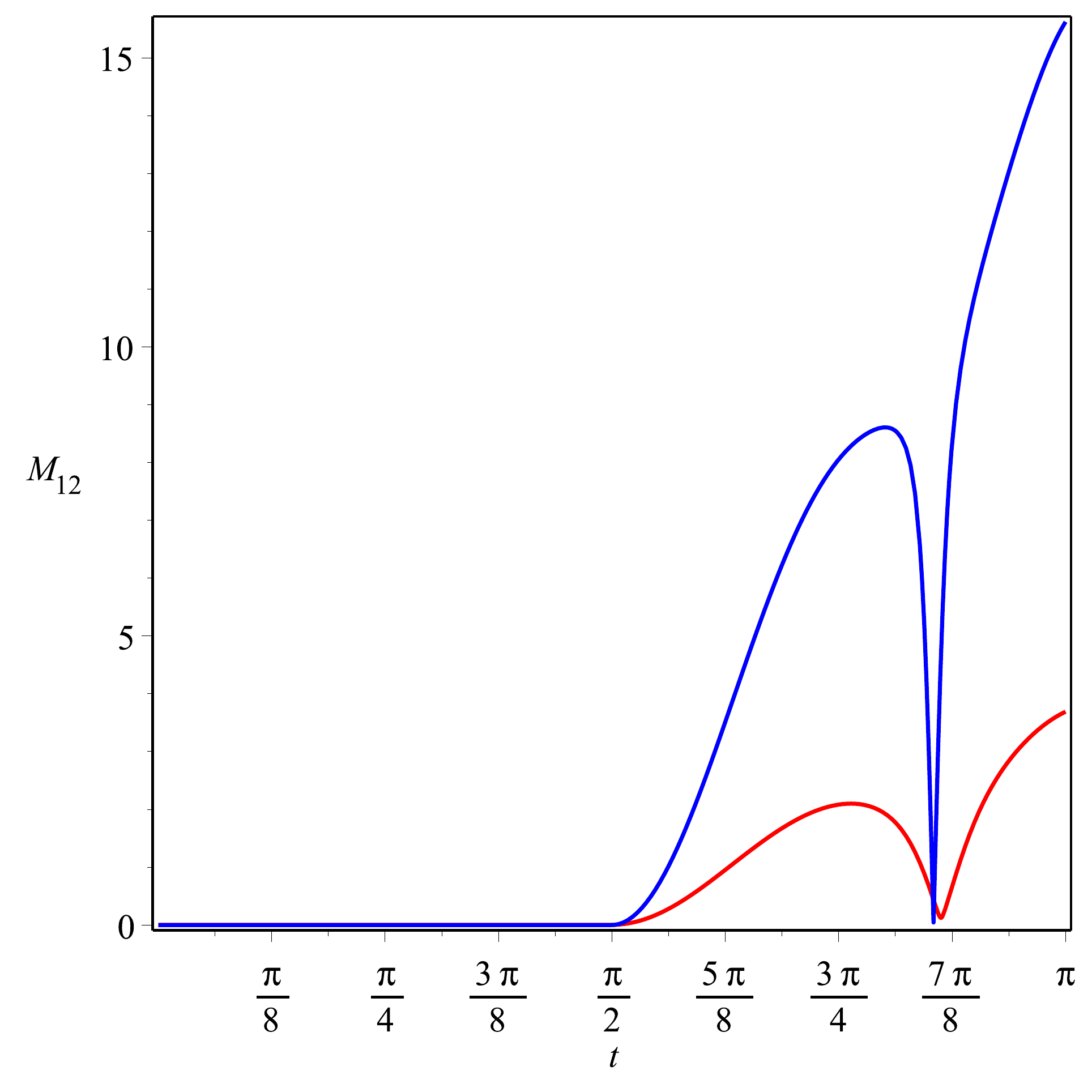}
	\captionof{figure}{\sf (color online)  Left panel presents the dynamics of entanglement quantified by logarithmic negativity $ E_{\mathcal{N}} (t)$. Right panel shows the geometric average of photon excitation's $M_{12}(t)= (\langle N_{1} \rangle \langle N_{2} \rangle)^{\frac{1}{2}} $.  The configuartions are taken  $\omega_{0}^{2}=1.01$, (blue line) for
		$( \epsilon=0.99, J_{0}=0.01)$
		and   (red line) for $( \epsilon=0.9, J_{0}=0.1)$. \label{heighquench}	  }
\end{figure}

 In {\color{blue} Figure} \ref{entanvsphon}, we plot  the link between the  entanglement and photon excitation's when the oscillators are resonant    $\omega_{1}^{2}=\omega_{2}^{2}=1.01$ ($\epsilon=0$) for three values of coupling $ J_{0}=0.1, 0.2, 0.3 $. {We observe that entanglement and vaccum excitation's are freezed during the half period $ [0,\frac{\pi}{2}] $. This is due to the fact that dilatation functions (\ref{erm}-\ref{Erm}) still constant in this interval. For $t>\frac{\pi}{2}$, the both quantities  exhibit the same topological behavior  ($\langle N \rangle\propto E_{\mathcal{N}}$). 

\begin{figure}[H]
	\centering
	\includegraphics[width=6cm, height=6cm]{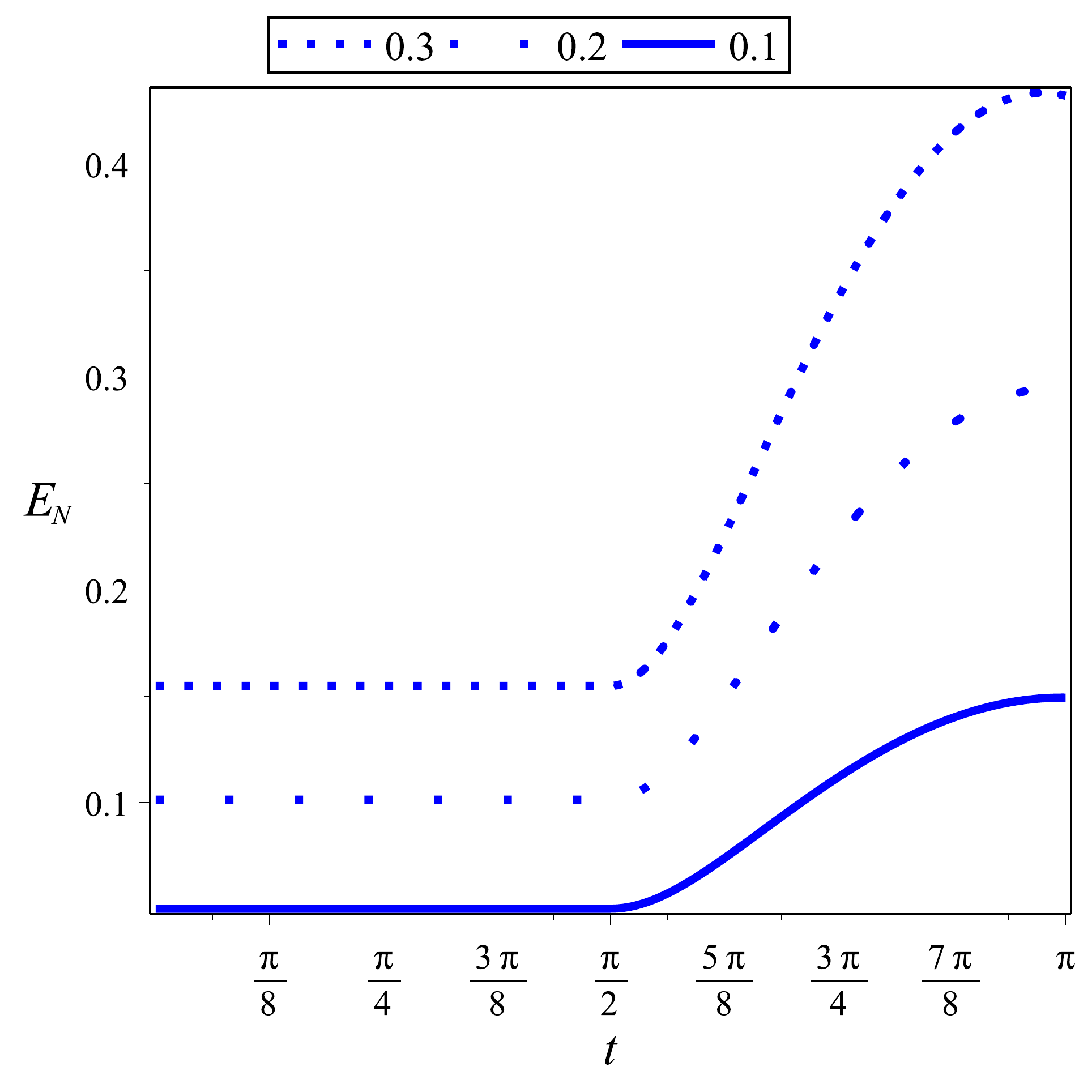}\ \ \ \ \ \
	\includegraphics[width=6.3cm, height=6cm]{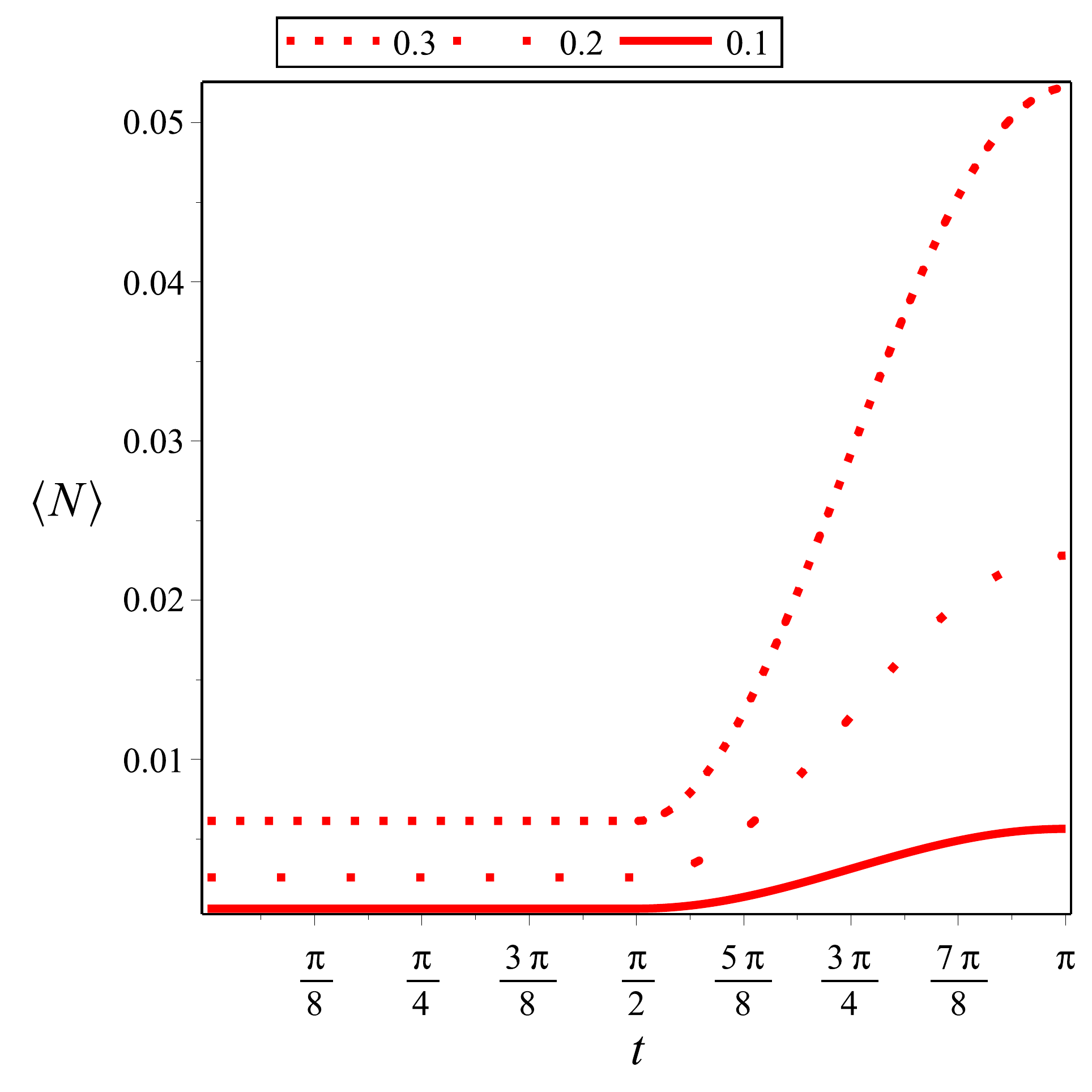}
	\captionof{figure}{\sf (color online)  Left panel presents the dynamics of entanglement quantified by Logarithmic negativity $ E_{\mathcal{N}} $ and  right panel shows the photon excitation's $ \langle N_{1} \rangle=\langle N_{2} \rangle $$=\langle N \rangle $ for the parameters $ \epsilon=0$, $\omega_{0}^{2}=1.01$ and $ J_{0}=0.1,0.2 $, $ 0.3 $.
		\label{entanvsphon} 	  }
\end{figure}

  In  {\color{blue} Figure} \ref{heighquench}, we investigate the effects of non-resonance on the geometric average of excitation's $ M_{12}(t)=\left(\left\langle N_{1}(t)\right\rangle \left\langle N_{2}(t)\right\rangle  \right)^{\frac{1}{2}}$ and entanglement.  When $ t\in [0,\frac{\pi}{2}] $ a tiny and constant amount of excitation's and entanglement are detected, but since $t>\frac{\pi}{2}$ the excitation's dramatically increase compared to the entanglement generation.  We mention, although the
   excitation's are important  the oscillators still weekly entangled and  the maximum of entanglement is obtained when the excitation's are more hierarchical (i.e.$ |\left\langle N_{2}(t)\right\rangle)-\left\langle N_{1}(t)\right\rangle| $ reaches its maximal value). To conclude,  the extinction of entanglement requires  the condition $ J_{0}=0 $, but that of excitations requires  resonance together with $ J_{0}=0 $. The photon excitation's generate important amounts of  entanglement when oscillators are resonant $ (\epsilon=0)$ and strongly coupled.
   This is similar to what has been found by considering  time-independent coupled oscillators \cite{R7}. It turns out this phenomenon can be seen as
 the Casimir effect  because   the virtual photon
 exchange plays a vital role in mediating the
 coupling between oscillators \cite{R16}.   
 Then they generate entanglement,  but their contribution is perhaps limited by their quantum destructive interference, which becomes more important when the system moves away from the resonance $(\epsilon\longrightarrow \omega_{0}^{2})$.}

\begin{figure}[H] 
	\centering
	\includegraphics[width=6cm, height=6cm]{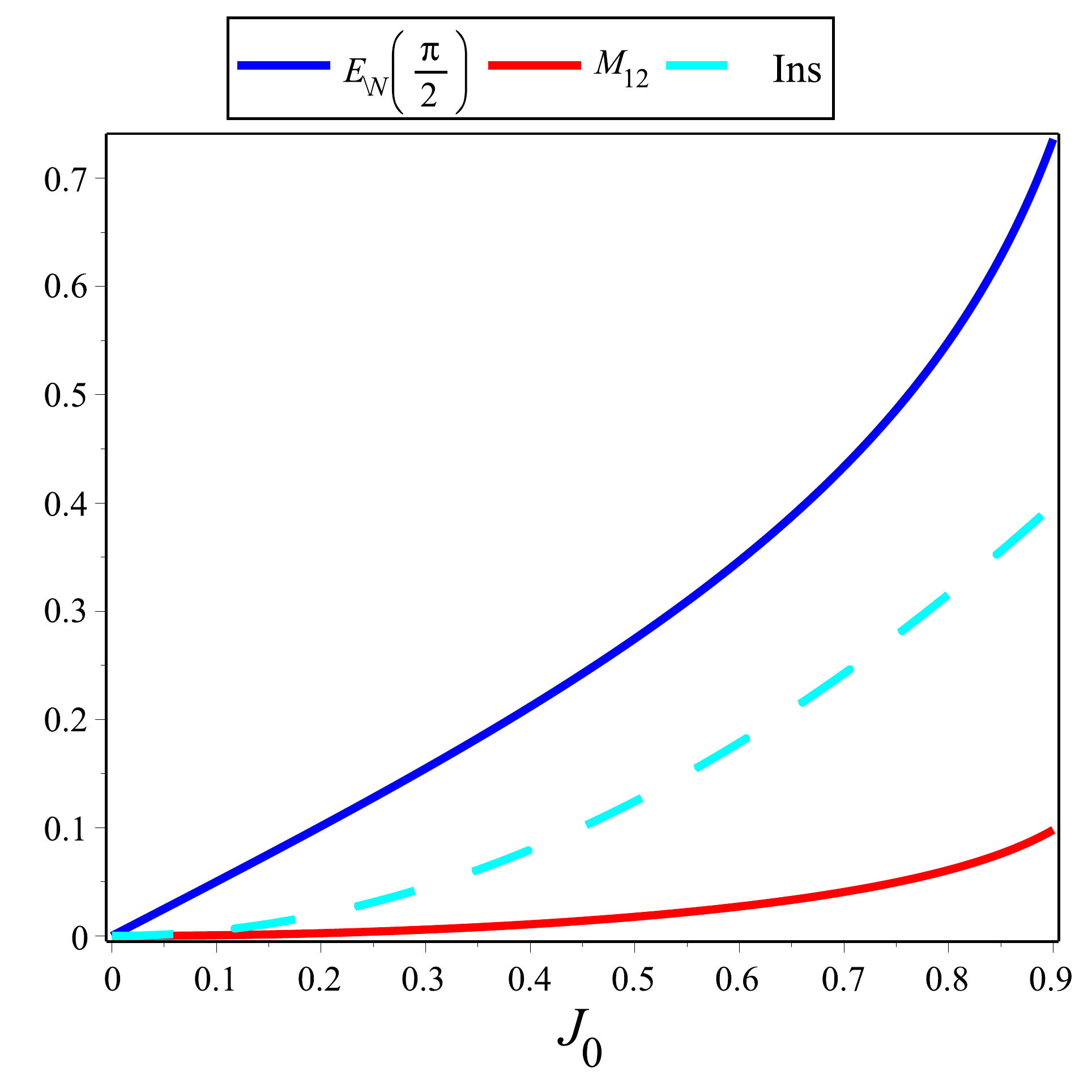}\ \ \ \ \
	\includegraphics[width=6cm, height=6cm]{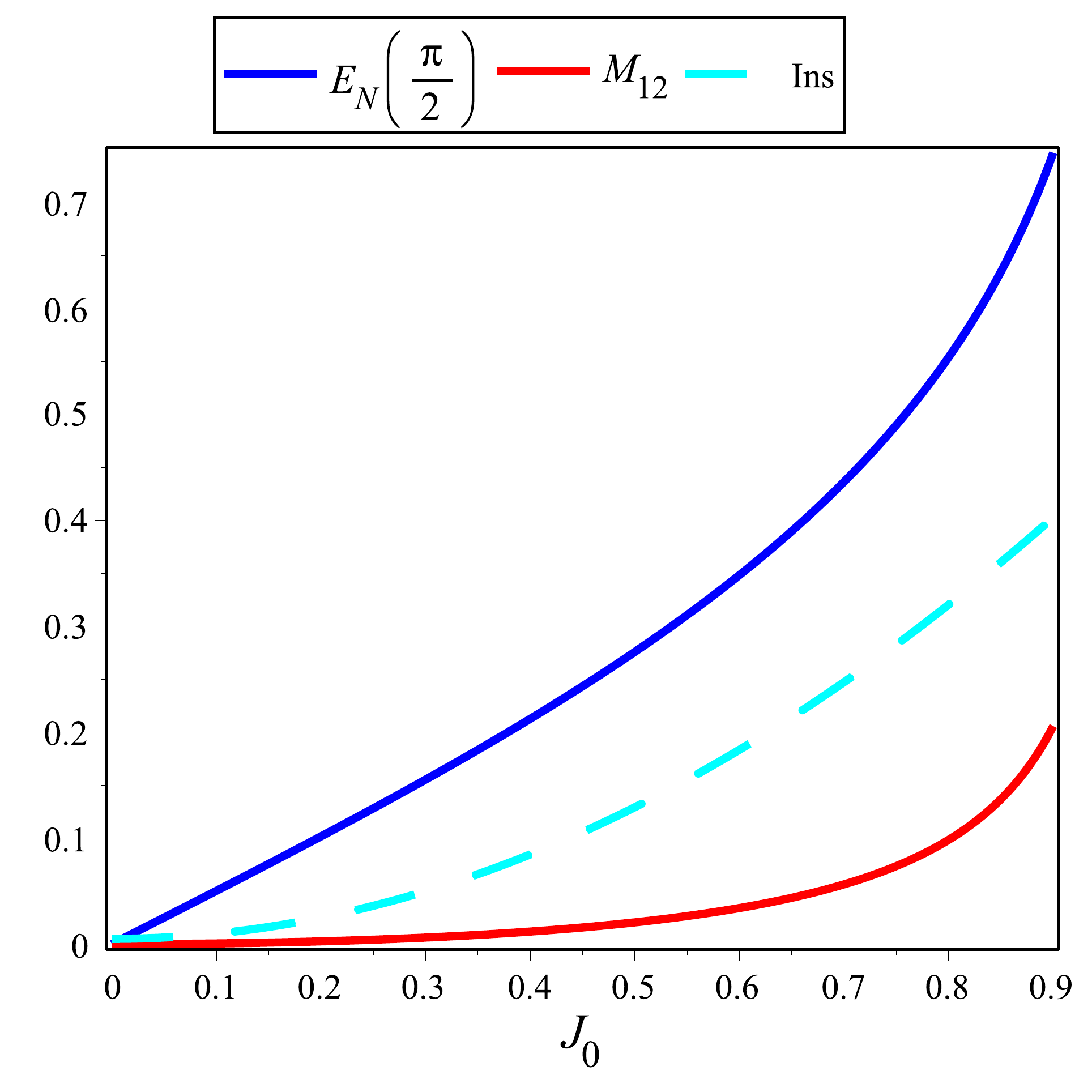}
	\captionof{figure}{\sf (color online)  Effects of  coupling   on entanglement $E_{\mathcal{N}}(\frac{\pi}{2})$,  geometric average $M_{12}=\left( \langle N_{1}\rangle \langle N_{2}\rangle\right)^{\frac{1}{2}}$ and $ Ins = \Lambda(\Omega_{1},\Omega_{2})-1 $, left panel: $\epsilon=0$ and 
		right panel:$\epsilon=0.1$ with  $\omega_{0}^{2}=1.01$. \label{entanglement}	  }
\end{figure}

In {\color{blue} Figure} \ref{entanglement}, we show the effects of coupling on entanglement $E_{\mathcal{N}}(t)$, geometric average of photon $ M_{12}(t)=\left( \langle N_{1}(t)\rangle \langle N_{2}(t)\rangle\right)^{\frac{1}{2}}$,  $ (t\in[0,\pi/2])$, and  we define $ Ins:=\Lambda(\Omega_{1},\Omega_{2})-1 $ as  geometric quantifier of instability (\ref{stabb}). We observe that the three quantities exhibit the same behavior with respect to coupling $ J_{0} $, they are monotonically increasing,  which entails the importance of stong coupling physics \cite{R7}. It is worthy to note that $ Ins(J_{0}=0,\epsilon=0.1)\sim 0.005 $ (oscillators are unstable), $ E_{\mathcal{N}}=0 $ and $M_{12}(t)\neq 0 $, similarly, $ Ins(J_{0}=0,\epsilon=0)\sim -1.2. 10^{-4} $ (oscillators are stable), $ E_{\mathcal{N}}=0 $ and $M_{12}(t)= 0 $. Finally,  those similarities show that classical instabilities affect the generation of  excitation's and  hence entanglement between oscillators. \\

\section{Conclusion}

We have studied two non resonant coupled harmonic oscillators connected by a periodically pumped  coupling $J(t)=J_{0}\times \Theta(t)$ and  frequencies. We have solved exactly the Schr\"{o}dinger dynamics, which leads to Ermakov differential equations. After a suitable transformations, we have showed that they are equivalent to classical counterparts of the decoupled  quantum Hamiltonian.  By studying the instability of the classical analog and computing the photon excitation's in the vacuum state, we have found that the excitation's will be created if the classical oscillators are unstable.

{We have studied the dynamics of entanglement by computing logarithmic negativity together with  photon excitation's beyond resonance by using phase space picture. We have analyzed the effects of coupling and quench amplitude on entanglement dynamics and photon excitation's. Consequently, it was found that photon excitation's present the same behavior when the oscillators are resonant and strongly coupled $ J_{0}\sim \omega^{2}_{1,2} $, and when the oscillators are  non resonant and weakly  coupled  the  photon's excitation's have a tiny contribution on entanglement generation. We have showed also that  extinction of excitation's entails the suppression of entanglement.} However, it  does not imply 
necessarily the  suppression of excitation's  if the oscillators are separated. This allows us to conclude that the excitation's generate and maintain entanglement.


\begin{thebibliography}{99}
\bibitem{R01}	A. Einstein, B. Podolsky, and N. Rosen, Phys. Rev. 47, 777
	(1935).
	\bibitem{R02}	J. S. Bell, Physics 1, 195 (1964). 
	\bibitem{R03}J. C. Gonzalez-Henao, E. Pugliese, S. Euzzor, S. F. Abdalah, R. Meucci and J. A. Roversi, Scientific Reports 5, 13152 (2015).
	\bibitem{R04} J. C. Gonzalez-Henao, E. Pugliese, S. Euzzor, R. Meucci, J. A. Roversi and F. T. Arecchi, 
	Scientific Reports 7, 9957 (2017). 
	\bibitem{R05} T. Figueiredo Roque and J. A. Roversi, Phys. Rev. A 88, 032114 (2013).
	\bibitem{R7} J.-Y. Zhou, Y.-H. Zhou, X.-L. Yin, J.-F.  Huang and J.-Q. Liao, Scientific Reports 10, 12557 (2020).

		\bibitem{R06}  E. Meissner, 
	Schweizer Bauzeitung 72, 95 (1918).
	\bibitem{R10} J. A. Richards, {\em Analysis of periodically time-varying systems} (Springer-Verlag, Berlin, 1983).
	\bibitem{R15} A. A. Burov and V. I. Nikonov, Int. J. Non-Linear Mech. 110, 26 (2019).
	\bibitem{R9} E. Pinney,   Proc. Am. Math. Soc. 1, 681, (1950).
		\bibitem{R2}  D. X. Macedo  and I. Guedes, J. Math. Phys. 53, 052101 (2012).
		\bibitem{R17} S. Menouar, M. Maamache and J. R. Choi, Physica Scripta 82, 6 (2010).
		\bibitem {R5} H. R. Lewis  and W. B. Riesenfeld, J. Math. Phys. 10, 1458 (1969).
		\bibitem{korsch98} K. E. Thylwe and H. J. Korsch,
		J. Phys. A 31, L279–L285 (1998).
		
		
		
		
		
		\bibitem{R11}Xi Chen, A. Ruschhaupt, S. Schmidt, A. del Campo, D. Guéry-Odelin and J. G. Muga, Phys.
		Rev. Lett. 104, 063002 (2010).
		\bibitem{R12} A. Tobalina, E. Torrontegui, I. Lizuain, M. Palmero and J. G. Muga, Phys. Rev. A 102, 063112 (2020).
		\bibitem{R1} G. Adesso and F. Illuminati, Phys. Rev. A 72, 032334 (2005).  
		\bibitem{R6} G. Adesso, A. Serafini and F. Illuminati, Phys. Rev. A 73, 032345 (2006).
		
		\bibitem{R13} Y. S. Kim and M. E. Noz, {\em Phase Space Picture of Quantum Mechanics} (World Scientific,
		Singapore, 1991).
		\bibitem{R14}M. A. Lohe,
		J. Phys. A: Math. Theor. 42,  035307 (2009).



\bibitem{R8} K. H. Yeon, H. J. Kim, C. I. Um, T. F. George and L. N. Pandey, Phys. Rev. A 50, 1035 (1994).


\bibitem{R16} D. L. Andrews and D. S. Bradshaw, Ann. Phys.  526, 173 (2014).



\end{thebibliography}
\end{document}